\documentclass[a4paper,11pt]{article}
\pdfoutput=1
\usepackage{jheppub}

\usepackage{tensor}

\usepackage[T1]{fontenc}
\usepackage[utf8]{inputenc}

\newcommand{\pd}{\partial}

\newcommand{\vphi}{\varphi}
\newcommand{\eps}{\varepsilon}

\newcommand{\OOO}[1]{\mathcal{O}\left(#1\right)}
\newcommand{\LL}{\mathcal{L}}
\newcommand{\CL}{\mathcal{L}}
\newcommand{\CE}{\mathcal{E}}
\newcommand{\CR}{\mathcal{R}}
\newcommand{\CB}{\mathcal{B}}

\newcommand{\ua}{{\underline{a}}}
\newcommand{\ub}{{\underline{b}}}

\newcommand{\spa}{\, , \qquad}

\newcommand{\inv}{^{-1}}

\newcommand{\mB}{\mathcal{B}}

\def\mM{\mathcal{M}}
\def\mN{\mathcal{N}}

\makeatletter
\g@addto@macro\bfseries{\boldmath}
\makeatother

\title{Relating non-relativistic string theories}

\author[a]{Troels Harmark}
\author[b]{Jelle Hartong}
\author[ac]{Lorenzo Menculini}
\author[ad]{Niels A. Obers}
\author[a]{Gerben Oling}

\affiliation[a]{The Niels Bohr Institute, University of Copenhagen,\\
Blegdamsvej 17, DK-2100 Copenhagen {\O}, Denmark}
\affiliation[b]{School of Mathematics and Maxwell Institute for Mathematical Sciences, University of Edinburgh,\\
Peter Guthrie Tait Road, Edinburgh EH9 3FD, UK}
\affiliation[c]{Dipartimento di Fisica e Geologia, Universit\`a di Perugia,\\
I.N.F.N. Sezione di Perugia,\\
Via Pascoli, I-06123 Perugia, Italy}
\affiliation[d]{Nordita, KTH Royal Institute of Technology and Stockholm University,\\
Roslagstullsbacken 23, SE-106 91 Stockholm, Sweden}

\emailAdd{harmark@nbi.ku.dk}
\emailAdd{jelle.hartong@ed.ac.uk}
\emailAdd{lorenzo.menculini@nbi.ku.dk}
\emailAdd{obers@nbi.ku.dk}
\emailAdd{gerben.oling@nbi.ku.dk}

\abstract{%
Non-relativistic string theories promise to provide simpler theories of quantum gravity as well as tractable limits of the AdS/CFT correspondence.
However, several apparently distinct non-relativistic string theories have been constructed.
In particular, one approach is to reduce a relativistic string along a null isometry in target space.
Another method is to perform an appropriate large speed of light expansion of a relativistic string.
Both of the resulting non-relativistic string theories only have a well-defined spectrum if they have nonzero winding along a longitudinal spatial direction.
In the presence of a Kalb--Ramond field, we show that these theories are equivalent provided the latter direction is an isometry.
Finally, we consider a further limit of non-relativistic string theory that has proven useful in the context of AdS/CFT (related to Spin Matrix Theory).
In that case, the worldsheet theory itself becomes non-relativistic and the dilaton coupling vanishes.
}

\begin{document}
\maketitle
\flushbottom

\section{Introduction}
\label{sec:intro}
Recent years have seen a surge of interest in applying Newton--Cartan (NC) geometry and its torsionful
generalization%
\footnote{Torsional Newton--Cartan (TNC) geometry is a torsionful version of standard NC geometry,  as originally introduced by Cartan (see \cite{Andringa:2010it} for a modern perspective and earlier references)
to geometrize Newtonian gravity. The torsionful generalization of NC geometry allows for a non-exact clock form and was first observed as the boundary geometry in the context of Lifshitz holography
\cite{Christensen:2013lma,Christensen:2013rfa,Hartong:2014pma}. It also plays an important role as the background geometry  that non-relativistic field theories couple to \cite{Jensen:2014aia,Hartong:2014oma}, with
the fractional quantum Hall effect \cite{Son:2013rqa,Geracie:2014nka,Gromov:2014vla,Geracie:2015dea} as one of its first applications
in that context.} to the study of non-relativistic aspects of field theory, gravity, string theory and holography.
The development of NC geometry has established a framework  for a covariant formulation of physics in non-relativistic limits, $1/c$ expansions as well as null reductions, which has already generated numerous novel insights.
Among the most fundamental of these is that applying NC geometry (and closely related avatars) to non-relativistic string theory
\cite{Andringa:2012uz,Harmark:2017rpg,Kluson:2018egd,Bergshoeff:2018yvt,Kluson:2018grx,Harmark:2018cdl,
Kluson:2019ifd,Gomis:2019zyu,Gallegos:2019icg} promises to open a window into non-relativistic quantum gravity.

Formulating a theory of non-relativistic quantum gravity using string theory would provide  a third road from which one can approach relativistic quantum gravity, supplementing the two more conventional roads
that start from relativistic quantum field theory and general relativity, respectively.
In fact, already at the classical level, there appear to be different types of non-relativistic theories of gravity
\cite{Hartong:2015zia,Bergshoeff:2015uaa,Afshar:2015aku,
Bergshoeff:2016lwr,Hartong:2016yrf,VandenBleeken:2017rij,Bergshoeff:2017dqq,Hartong:2017bwq,Aviles:2018jzw,Hansen:2018ofj,
Bergshoeff:2018vfn,Cariglia:2018hyr,Matulich:2019cdo,VandenBleeken:2019gqa,Hansen:2019vqf,Bergshoeff:2019ctr,deAzcarraga:2019mdn,Hansen:2019svu,Concha:2019lhn,Penafiel:2019czp}
depending on the specifics of the local non-relativistic spacetime symmetry algebra. In particular, the non-relativistic
limit of General Relativity itself turns out to be much richer \cite{VandenBleeken:2017rij,Hansen:2018ofj,VandenBleeken:2019gqa,Hansen:2019vqf,Hansen:2019svu,Hansen:2019}
than just Newtonian gravity.  All this raises the question whether there exist non-relativistic theories of gravity that admit
a UV completion in terms of a non-relativistic string theory.
For the purposes of this paper, we will mean by `non-relativistic string theory' the generalization of the Gomis--Ooguri non-relativistic action~\cite{Gomis:2000bd,Danielsson:2000gi} to arbitrary backgrounds.

So far, two types of non-relativistic string theories have been introduced, namely strings on
torsional Newton--Cartan  geometry \cite{Harmark:2017rpg,Harmark:2018cdl} and strings
on string Newton--Cartan  geometry \cite{Andringa:2012uz,Bergshoeff:2018yvt}.
Moreover, beta functions for these theories have recently been studied  in
\cite{Gomis:2019zyu} for the latter case and in \cite{Gallegos:2019icg} for the former.
Indeed, in line with their proposed relation to non-relativistic theories of gravity, the beta functions of these strings lead to dynamical equations  of the corresponding non-relativistic gravity theories.
The primary aim of the present work is to show that these two seemingly distinct non-relativistic string theories can in fact be mapped onto each other.

There are  many additional motivations to study non-relativistic string theory.  First of all it is a limit of string theory as we know it, and hence interesting in its own right, while at the same time some of its features could turn out to be simpler as compared to relativistic string theory.
Furthermore, non-relativistic string theory appears to be related to limits of the AdS/CFT correspondence, as shown in particular in  \cite{Harmark:2017rpg,Harmark:2018cdl} by taking a further non-relativistic limit on the worldsheet. In this setting, the connection with non-relativistic geometry elucidates the dual space-time formulation of Spin Matrix Theory (SMT)
\cite{Harmark:2008gm,Harmark:2014mpa}. SMT is a quantum mechanical theory obtained by considering near-BPS limits of AdS${}_5$/CFT${}_4$, which reduces to tractable sectors in which quantum gravity effects are potentially easier to compute.  Interestingly, the class of non-relativistic worldsheet sigma models that was found in  \cite{Harmark:2018cdl}
exhibits an infinite-dimensional symmetry, namely the two-dimensional Galilean Conformal Algebra.
The results of \cite{Harmark:2018cdl} also lead to novel interpretations of some classical results related to integrable 2D field theories. For example, the continuum limit of Heisenberg spin chains, described by the Landau-Lifshitz model, can be viewed as strings with non-relativistic worldsheets.
Finally, interesting connections between doubled field theory and certain non-relativistic strings have been found \cite{Ko:2015rha,Morand:2017fnv,Berman:2019izh}, which open another arena of connections between NC geometry and doubled (or generalized) geometries.

Non-relativistic string theory has a rich history, starting with the original papers \cite{Gomis:2000bd,Danielsson:2000gi}
 (see also \cite{Kruczenski:2003gt,Gomis:2005pg,Bagchi:2009my}) in which a Galilean invariant closed string theory was constructed with excitations satisfying a non-relativistic dispersion relation.
Following recent advances in understanding non-relativistic geometry,
two versions of non-relativistic string theory have been proposed. In one approach, null reduction was used to obtain a Nambu--Goto (NG) action~\cite{ Harmark:2017rpg} and corresponding Polyakov-type action~\cite{Harmark:2018cdl} for non-relativistic strings on torsional Newton--Cartan (TNC)  geometry. More precisely, the target space geometry is TNC geometry
plus an extra periodic target space direction on which there is a non-zero string winding.
Classically, the embedding coordinate along this direction is locally pure gauge, but this is not true globally due to the winding.
For practical purposes
we therefore refer to the target space as TNC geometry.
Following another direction, using an appropriate large-$c$ expansion an action for strings on a ``stringy'' Newton--Cartan geometry was proposed in
 \cite{Andringa:2012uz} and further developed in \cite{Bergshoeff:2018yvt}. We will denote the latter in the following as strings on  string Newton--Cartan (SNC) geometry.  Both of these non-relativistic string theories have  the property that they reduce to the Gomis-Ooguri action \cite{Gomis:2000bd} on the corresponding flat non-relativistic spacetime.

In this paper, we will demonstrate that (modulo an assumption stated below) these two theories can be related to each other.%
\footnote{Note that in Ref.~\cite{Harmark:2018cdl}  a map was given that related certain sectors of these two non-relativistic string actions. As already pointed out there, this map was not complete since it did not include the
NS-NS $B$-field. As we will see in the present work, including this background field turns out to be crucial.}
In particular, we will show that closed  string theory with a TNC target space accompanied by a corresponding NS-NS sector can be identified with SNC string theory, which involves two distinct longitudinal directions in the target space. The identification holds provided the compact spatial longitudinal direction of the SNC target space is an isometry.
An important element in the derivation is that the SNC action admits a field redefinition freedom,
which turns out to also reproduce certain symmetries coming from the underlying SNC algebra.
We also obtain the extension of the non-relativistic worldsheet theory of Ref.~\cite{Harmark:2018cdl} that includes
the NS-NS sector and dilaton, by considering a further limit of our general TNC action. In that case, we observe
that the worldsheet dilaton coupling term vanishes.

\subsection*{Summary and outline}
In Section \ref{sec:null-red} we first extend the derivation  of the Polyakov-type action for TNC strings \cite{Harmark:2017rpg} to include the $B$-field (see also \cite{Gallegos:2019icg}) and dilaton. This is done via a null reduction of the relativistic Polyakov action, fixing
the momentum along the null reduction and going to a dual formulation in which this momentum is exchanged for
a fixed winding along a dual compact direction.
We also present two alternative forms of this action, including the
generalization of coupling the full NS-NS sector to the Nambu--Goto type action of Ref.~\cite{Harmark:2017rpg}.

Section \ref{sec:snc-tnc-equivalence} is the main part of the paper, where we compare to the SNC action of \cite{Bergshoeff:2018yvt}.
We introduce this action by rederiving  it (closely paralleling the original work \cite{Andringa:2012uz})
from a large-$c$ expansion of the relativistic Nambu--Goto action coupled to the NS-NS sector in which two longitudinal directions are singled out among the target space directions.  This derivation will highlight two important points:
\begin{itemize}
  \item
    a field redefinition symmetry, which will be crucial to show the equivalence of the SNC action to the TNC action obtained in Section~\ref{sec:null-red},
  \item
    including the NS-NS sector lifts a constraint (closely
related to the foliation constraint of \cite{Bergshoeff:2018yvt}) that is present at the classical level in the absence
of the Kalb--Ramond field.
\end{itemize}
The direct map between the TNC and SNC action, which holds in the case
that the compact spatial longitudinal direction is an isometry, is then presented in Section~\ref{ssec:map-from-snc-with-isometry-to-tnc}.

In Section \ref{sec:nr-ws-limit} we examine the effect of including the NS-NS sector and dilaton on the class of non-relativistic worldsheet  sigma-models obtained in \cite{Harmark:2017rpg,Harmark:2018cdl} by taking a second limit. After presenting
the resulting sigma model we discuss the dilaton term in more detail and observe a suggestive interplay between
the scaling of this term and the allowed worldsheet geometries.

A number of details are included in appendices. In Appendix \ref{app:conventions} we collect our
conventions for the worldsheet vielbeine and list some useful identities.
Furthermore, in Appendix \ref{app:SNC-TNC-symmetries} we first review SNC geometry and its associated symmetry algebra.
We also show how this algebra can be obtained from an expansion of the Poincar\'e  algebra and indicate how the $B$-field could be included in the gauging procedure.
Finally, we give details on the identifications of symmetries between TNC and SNC strings.

\vspace{\baselineskip}
\emph{Note added:} While this paper was being completed, an analysis of the beta functions of the TNC string including NS-NS background fields appeared~\cite{Gallegos:2019icg} where the action in~\eqref{eq:tnc-ns-string-action-vielbein} and its $U(1)$ symmetries were derived independently.

\section{Null reduction of relativistic strings in NS-NS background}
\label{sec:null-red}

The purpose of this section is to extend the derivation of the action for non-relativistic strings from a null reduction%
\footnote{For clarity, we note that this is not ordinary Kaluza-Klein reduction, but a dimensional
reduction in which the momentum along the null direction $u$ is fixed.}
in~\cite{Harmark:2018cdl} to include NS-NS background fields.
To do this, we start from the full NS-NS action in a relativistic, Lorentzian spacetime and reduce along a null isometry.
The TNC string action is constructed by implementing the conservation of the string momentum along the null isometry direction using Lagrange multipliers and by then going to a dual formulation.
In this process, the total momentum along the null direction is fixed and can be exchanged for a fixed winding of the string along a compact dual direction.

\subsubsection*{Brief review of TNC geometry}
As is well known, torsional Newton--Cartan (TNC) geometry in $d+1$ spacetime dimensions can be obtained from a $(d+2)$-dimensional Lorentzian geometry with a non-compact null isometry.
We choose coordinates such that this isometry is generated by~$\pd_u$.
Then we can split $(d+2)$-dimensional coordinates $x^\mM=0,1,\ldots,d+1$ as $x^\mM=(u,x^\mu)$, where $x^\mu$ are $(d+1)$-dimensional spacetime coordinates.
Without loss of generality, we can then write any $(d+2)$-dimensional metric $G_{\mM\mN}$ as
\begin{equation}
  \label{eq:metric-decomposition}
  ds^2
    = G_{\mM\mN} dx^{\mM} dx^{\mN}
    = 2\tau(du-m) + h_{\mu\nu} dx^\mu dx^\nu.
\end{equation}
In this decomposition, all components of the metric are independent of $u$.
The one-forms $\tau=\tau_\mu dx^\mu$ and $m=m_\mu dx^\mu$, together with the symmetric two-tensor $h_{\mu\nu}dx^\mu dx^\nu$, define a torsional Newton--Cartan structure.
Distances in space and time are measured using $h^{\mu\nu}$ and $\tau_\mu$, respectively, where the associated projective inverses $v^\mu$ and $h^{\mu\nu}$ satisfy
\begin{equation}
  \label{eq:nc-metrics-inverse-relations}
  v^\mu \tau_\mu = -1\, ,
    \quad
  v^\mu h_{\mu\nu} = 0\, ,
    \quad
  \tau_\mu h^{\mu\nu} = 0\, ,
    \quad
  h_{\mu\rho} h^{\rho\nu} = \tau_\mu v^\nu + \delta_\mu^\nu\, .
\end{equation}
We can introduce vielbeine $\tensor{\CE}{_\mu^a}$ for $h_{\mu\nu}$ as
\begin{equation}
  \label{eq:spatial-metric-vielbeine}
  h_{\mu\nu}=\CE_\mu{}^a \CE_\nu{}^b \delta_{ab} \spa  a,b = 1 \ldots d\, .
\end{equation}
The TNC fields transform as tensors under diffeomorphisms $\xi^\mu$ and exhibit furthermore a set of
local symmetries corresponding to Galilean (or Milne) boosts $\lambda_\mu$
and a $U(1)$ gauge transformation $\sigma$ associated with mass conservation,
\begin{equation}
\begin{gathered}
\delta\tau_\mu=\mathcal{L}_\xi\tau_\mu \spa\delta h_{\mu \nu} =\mathcal{L}_\xi h_{\mu \nu} +\lambda_\mu \tau_\nu+
\lambda_\nu \tau_\mu \,, \\
\delta m_\mu=\mathcal{L}_\xi m_\mu+\lambda_\mu+\partial_\mu\sigma\, ,  \label{eq:typeITNC}
\end{gathered}
\end{equation}
where $\mathcal{L}_\xi$ denotes the Lie derivative along $\xi^\mu$.

\subsection{NS-NS action for TNC strings from null reduction}
\label{ssec:nsns-tnc-action-from-null-reduction}
Our starting point is the usual relativistic bosonic string action with a $(d+2)$-dimensional target space, parametrized using coordinates $x^\mM$, including a Kalb--Ramond two-form gauge field $\CB_{\mM\mN}$ and a dilaton $\phi$,
\begin{equation}
  \label{eq:higher-dim-ns-string-action}
  \begin{split}
    S[\gamma_{\alpha\beta},X^\mM]
    &= - \frac{T}{2} \int_\Sigma d^2\sigma
      \left(
        \sqrt{|\gamma|}\gamma^{\alpha\beta} \pd_\alpha X^\mM \pd_\beta X^\mN G_{\mM\mN}
        + \eps^{\alpha\beta} \pd_\alpha X^\mM \pd_\beta X^\mN \mB_{\mM\mN}
      \right)
    \\
    &{}\qquad
      + \frac{1}{4\pi} \int_\Sigma d^2\sigma \sqrt{|\gamma|} R^{(2)} \phi\,.
  \end{split}
\end{equation}
Here, $\gamma^{\alpha\beta}$ is the Lorentzian metric on the string worldsheet $\Sigma$, which is parametrized using the coordinates~$\sigma^\alpha = (\sigma^0,\sigma^1)$.
We restrict ourselves to closed strings, with $\sigma^1 \sim \sigma^1 + 2\pi$.
Furthermore, we set $\eps^{01}=+1$.
Following the decomposition of the $(d+2)$-dimensional metric~$G_{\mM\mN}$ in~\eqref{eq:metric-decomposition}, we can split $\mB_{\mM\mN}$ as follows:
\begin{equation}
  \label{eq:bfield-decomposition}
    \mB_{\mu\nu}\, ,
      \qquad
    b_\mu = \mB_{u\mu}\, .
\end{equation}

Following the derivation in~\cite{Harmark:2018cdl}, we want to restrict the dynamics of the strings along the $u$-direction to construct an action for strings on the $(d+1)$-dimensional TNC spacetime.
This will result in exchanging the fixed momentum along the (non-compact) $u$-direction for a fixed winding along a compact dual direction.
Since we consider a fixed momentum sector, this does not correspond to the DLCQ procedure.%
\footnote{Considering a fixed momentum sector is similar to what happens in the particle case.
  The action
  that describes the coupling to TNC geometry of a non-relativistic particle of mass $m$ can be obtained from a null reduction of a massless relativistic particle by fixing the momentum along the (non-compact) null direction to be $m$
(see e.g. \cite{Festuccia:2016caf}).}

To find the worldsheet current $P_u^\alpha$ associated to the target space isometry $\pd_u$, we use the metric and $B$-field decompositions~\eqref{eq:metric-decomposition} and \eqref{eq:bfield-decomposition} to write
  \begin{align}
    \label{eq:ns-string-action-null-direction-split}
    S[\gamma_{\alpha\beta}, X^\mu, X^u]
      &= - \frac{T}{2} \int_\Sigma d^2\sigma
        \left(
          \sqrt{|\gamma|}\gamma^{\alpha\beta} \bar h_{\alpha\beta}
          + \eps^{\alpha\beta} \CB_{\alpha\beta}
        \right)
      \\
      &{}\qquad
      - T \int_\Sigma d^2\sigma
        \left(
          \sqrt{|\gamma|}\gamma^{\alpha\beta} \tau_\beta
          + \eps^{\alpha\beta}b_\beta
        \right)
        \pd_\alpha X^u
      + \frac{1}{4\pi} \int_\Sigma d^2\sigma \sqrt{|\gamma|} R^{(2)} \phi\, . \nonumber
  \end{align}
Here, we have introduced the Galilean (or Milne) boost-invariant tensor
\begin{equation}
  \bar h_{\mu\nu} = h_{\mu\nu} - \tau_\mu m_\nu - \tau_\nu m_\mu\, ,
\end{equation}
as well as the worldsheet pullbacks
\begin{equation}
  \bar h_{\alpha\beta} = \pd_\alpha X^\mu \pd_\beta X^\nu \bar h_{\mu\nu}\, ,
    \quad
  \tau_\alpha = \pd_\alpha X^\mu \tau_\mu\, ,
    \quad
  \CB_{\alpha\beta} = \pd_\alpha X^\mu \pd_\beta X^\nu \CB_{\mu\nu}\, ,
    \quad
  b_\alpha = \pd_\alpha X^\mu b_\mu\, .
\end{equation}
Note that the pullback only uses the $(d+1)$-dimensional embedding coordinates $X^\mu$.
From the form of the action in~\eqref{eq:ns-string-action-null-direction-split}, it is clear that the current $P_u^\alpha$ is given by
\begin{equation}
  \label{eq:u-momentum-current}
  P_u^\alpha
    =  \frac{\pd \LL}{\pd(\pd_\alpha X^u)}
    = - T \sqrt{|\gamma|} \gamma^{\alpha\beta}\tau_\beta - T\eps^{\alpha\beta} b_\beta\, .
\end{equation}

We then introduce an action where the conservation $\pd_\alpha P_u^\alpha=0$ is implemented by Lagrange multipliers, which allow us to define a dual direction to $u$.
To this end, consider the auxiliary action
\begin{align}
  \label{eq:tnc-ns-string-action-metric}
    S[\gamma_{\alpha\beta}, X^\mu, \eta, A_\alpha]
      &= - \frac{T}{2} \int_\Sigma d^2\sigma
        \left(
          \sqrt{|\gamma|}\gamma^{\alpha\beta} \bar h_{\alpha\beta}
          + \eps^{\alpha\beta} \mB_{\alpha\beta}
        \right)
      \\
      &{}\quad\nonumber
      - T \int_\Sigma d^2\sigma
        \left(
          \sqrt{|\gamma|}\gamma^{\alpha\beta} \tau_\beta
          + \eps^{\alpha\beta}(b_\beta
          + \pd_\beta \eta)
        \right)
        A_\alpha
      + \frac{1}{4\pi} \int_\Sigma d^2\sigma \sqrt{|\gamma|} R^{(2)} \phi \, .
\end{align}
In this action, we have exchanged $X^u$ for $\eta$ and $A_\alpha$.
It is then natural to interpret $A_\alpha$ as a Lagrange multiplier which imposes a condition on $\eta$ that we can solve in terms of $P_u^\alpha$,
\begin{equation}
  \label{eq:eta-u-current-relation}
 T d\eta
    =T \pd_\alpha \eta\, dx^\alpha
    = \eps_{\alpha\beta}P_u^\beta dx^\alpha\, .
\end{equation}
Since $d^2\eta=0$ this implies that $\pd_\alpha P_u^\alpha=0$.
We see that conservation of the $u$-translation current follows automatically from the constraint imposed by $A_\alpha$.
We denote the conserved total momentum in the $u$-direction by
\begin{equation}
  \label{eq:total-u-momentum}
  P
    = \oint P_u^0 d\sigma^1
    = - T \oint \left( \sqrt{|\gamma|}\gamma^{0\alpha} \tau_\alpha + b_1 \right) d\sigma^1\, .
\end{equation}

From the point of view of the $(d+2)$-dimensional target space parametrized by $x^\mM$, this means we consider strings with momentum $P$ in the $u$-direction.
Furthermore, because of the relation~\eqref{eq:eta-u-current-relation}, $P$ is equivalently encoded by the winding of the string in $\eta$.
It is therefore convenient to parametrize $\eta$ using
\begin{equation}
  \eta(\sigma^0,\sigma^1) = \frac{P}{2\pi T} \sigma^1 + \tilde\eta(\sigma^0,\sigma^1)\, .
\end{equation}
Here, $\tilde\eta$ is an arbitrary function which is periodic in $\sigma^1$.
For the reasons outlined above, when varying the action~\eqref{eq:tnc-ns-string-action-metric} with respect to $\eta$, only $\tilde\eta$ is dynamical and $P$ is fixed.
The resulting equations of motion then imply $dA=0$.

Consequently, we can write $A=d\chi$, where in general $\chi$ can have winding.
Since we want to identify $\chi=X^u$, we should require that $A$ is exact because the string cannot wind along the non-compact null direction $u$.
Then the term involving $\eta$ in the action~\eqref{eq:tnc-ns-string-action-metric} drops out because $A=dX^u$ is exact,
\begin{equation}
  T\int_\Sigma A\wedge  d\eta
  = T \int_\Sigma d\left(X^u d\eta\right)
  = 0\, .
\end{equation}
Hence, we see that the action~\eqref{eq:tnc-ns-string-action-metric} is \emph{equivalent} to the action~\eqref{eq:ns-string-action-null-direction-split} once we solve the equations of motion associated to $\eta$ and $A$, if we furthermore demand that $A=dX^u$ with no winding along $X^u$.

Conversely, we can also interpret the worldsheet scalar~$\eta$ as the embedding coordinate~$X^v$ of the string along a dual compact direction $v$.
Since there is winding in $\eta$, the $v$-direction must be compact.
Here, the radius of the $v$-direction is~$P/(2\pi mT)$ for winding number $m$.
From this perspective, the requirement that $A$ is exact implies that the total momentum of the string along $v$ must be zero:
\begin{equation}
P_v^\alpha= \frac{\pd \LL}{\pd(\pd_\alpha X^v)}=T \eps^{\alpha\beta} A_\beta\, ,
  \quad\implies\quad
P_v=\oint P_v^0 d\sigma^1=T \oint A_1 =0\, .
\end{equation}
In other words, interpreting $\eta=X^v$ as an embedding coordinate instead of $A=dX^u$ exchanges the fixed momentum along the $u$-direction for a fixed winding along the $v$-direction.

In the following, it will be useful to denote the two distinct $(d+2)$-dimensional spacetimes that result from these two different interpretations of the action~\eqref{eq:tnc-ns-string-action-metric} using the coordinates
\begin{equation}
  x^\mM = (u, x^\mu)\, ,
    \qquad
  x^M = (v, x^\mu)\, .
\end{equation}
We will now briefly discuss the action with $x^M$ target space.
After that, we return to the TNC string action~\eqref{eq:tnc-ns-string-action-metric} and discuss its symmetries.

\subsection{Alternative forms of TNC action}

\label{ssec:tnc-action-uplift-vielbein-form}
Viewing $\eta=X^v$ as an embedding coordinate, we can reinterpret the action~\eqref{eq:tnc-ns-string-action-metric} in terms of the target space parametrized by $x^M=(v,x^\mu)$.
By setting
\begin{gather}
  \bar\CB_{M N} = \CB_{M N} + m_M b_N - m_N b_M\, ,
    \qquad
  \CB_{v \mu}=0\, ,
    \qquad
  b_v=1\, ,
    \\
    \nonumber
    \tau_v=0 \spa
   \CE_v{}^a=0 \spa
   h_{vv}= 0 \spa
   h_{v\mu} =0 \spa
  m_v= 0 \spa
  A_\alpha = \hat A_\alpha + m_\alpha\, ,
\end{gather}
we can write the action~\eqref{eq:tnc-ns-string-action-metric} in terms of the $X^M$ embedding coordinates as
\begin{align}
\label{eq:S-uplifted}
    S[\gamma_{\alpha\beta}, X^M, A_\alpha]
      &= - \frac{T}{2} \int_\Sigma d^2\sigma
        \left(
          \sqrt{|\gamma|}\gamma^{\alpha\beta} h_{MN}
          + \eps^{\alpha\beta} \bar\CB_{MN}
        \right)\pd_\alpha X^M \pd_\beta X^N
      \\
      &{}\quad\nonumber
      - T \int_\Sigma d^2\sigma
        \left(
          \sqrt{|\gamma|}\gamma^{\alpha\beta} \tau_M
          + \eps^{\alpha\beta} b_M
        \right)
       \pd_\beta X^M \hat A_\alpha
      + \frac{1}{4\pi} \int_\Sigma d^2\sigma \sqrt{|\gamma|} R^{(2)} \phi\, .
\end{align}
In this way, we have absorbed the TNC one-form $m_\mu$ in $\bar\CB_{\mu v}$, the components of the $(d+2)$-dimensional $B$-field along the $v$-direction.
Furthermore, to compare with existing actions, it is convenient to parametrize the worldsheet metric using zweibeine $e^a = e_\alpha{}^a d\sigma^\alpha$ so that $\gamma_{\alpha\beta} = \eta_{ab} e_\alpha{}^a e_\beta{}^b$.
(Note that $a=0,1$ for worldsheet quantities.)
If we then reparametrize the Lagrange multipliers using
\begin{equation}
  \hat A_\alpha
    = \frac{1}{2} \left(\lambda_+ - \lambda_-\right) e_\alpha{}^0
      + \frac{1}{2} \left(\lambda_+ + \lambda_-\right) e_\alpha{}^1\, ,
\end{equation}
the $(d+2)$-dimensional target space action becomes
\begin{align}
\label{eq:tnc-reinterpr}
   S[\gamma_{\alpha\beta}, X^M, \lambda_{\pm}]=
&- \frac{T}{2} \int_\Sigma d^2\sigma
        \left(   e \eta^{ab} e^\alpha{}_a e^\beta{}_b h_{MN}
          + \eps^{\alpha\beta} \bar \CB_{M N}
        \right)\pd_\alpha X^M \pd_\beta X^N
      \\\nonumber
     & - \frac{T}{2}  \int_\Sigma d^2\sigma
        \left[
          \lambda_+ \eps^{\alpha\beta} e_\alpha{}^+
            \left(\tau_M +  b_M \right)
          + \lambda_- \eps^{\alpha\beta} e_\alpha{}^-
            \left(\tau_M - b_M\right)
        \right]\pd_\beta X^M
      \\\nonumber
      &+ \frac{1}{4\pi} \int_\Sigma d^2\sigma \sqrt{|\gamma|} R^{(2)} \phi\, .
\end{align}
Here, we have defined $e_\alpha{}^\pm = e_\alpha{}^0 \pm e_\alpha{}^1$.
See Appendix~\ref{app:conventions} for our conventions and useful properties.

In the above, we have discussed two ways in which the auxiliary action~\eqref{eq:tnc-ns-string-action-metric} can be interpreted in terms of a $(d+2)$-dimensional geometry.
We first showed how it can reproduce the original relativistic string action on a Lorentzian target space with null isometry in~\eqref{eq:ns-string-action-null-direction-split}, and also related it to the target space parametrized by $x^M$ in~\eqref{eq:tnc-reinterpr}.
Alternatively, we can interpret~\eqref{eq:tnc-ns-string-action-metric} as a string action with a $(d+1)$-dimensional TNC target space supplemented by the worldsheet fields $A_\alpha$ and $\eta$.

As before, we redefine
\begin{equation}
  A_\alpha
    = m_\alpha
      + \frac{1}{2} \left(\lambda_+ - \lambda_-\right) e_\alpha{}^0
      + \frac{1}{2} \left(\lambda_+ + \lambda_-\right) e_\alpha{}^1\, ,
\end{equation}
so that we can write the action~\eqref{eq:tnc-ns-string-action-metric} as
\begin{align}
  \label{eq:tnc-ns-string-action-vielbein}
    S[\gamma_{\alpha\beta}, X^\mu, \eta, \lambda_{\pm}]
      &= - \frac{T}{2} \int_\Sigma d^2\sigma
        \left(
          2\eps^{\alpha\beta} m_\alpha \left(\pd_\beta\eta + b_\beta\right)
          + e \eta^{ab} e^\alpha{}_a e^\beta{}_b h_{\alpha\beta}
          + \eps^{\alpha\beta} \CB_{\alpha\beta}
        \right)
      \\\nonumber
      &{}\quad
      - \frac{T}{2}  \int_\Sigma d^2\sigma
        \left(
          \lambda_+ \eps^{\alpha\beta} e_\alpha{}^+
            \left(\tau_\beta + \pd_\beta\eta + b_\beta\right)
          + \lambda_- \eps^{\alpha\beta} e_\alpha{}^-
            \left(\tau_\beta - \pd_\beta\eta - b_\beta\right)
        \right)
      \\ \nonumber
      &{}\quad
      + \frac{1}{4\pi} \int_\Sigma d^2\sigma \sqrt{|\gamma|} R^{(2)} \phi\, .
\end{align}
In this case, $\CB_{\alpha\beta}=\pd_\alpha X^\mu \pd_\beta X^\nu \CB_{\mu\nu}$ and we retain the TNC one-form $m_\mu$ as a target space variable.
We also note that the action
\eqref{eq:tnc-ns-string-action-vielbein} reduces to the Gomis-Ooguri non-relativistic string action on a flat
TNC target space \cite{Harmark:2018cdl}.

\subsubsection*{Nambu--Goto form of action}
Until now we have kept the Lagrange multipliers in the Lagrangian, but we can also integrate them out.
This will lead us to the Nambu--Goto formulation of the TNC string action.
The constraints enforced by $\lambda_\pm$ are solved by taking
\begin{equation}
\label{eq:constr-solution}
e_\alpha{}^\pm = h_\pm (\tau_\alpha\pm \pd_\alpha \eta\pm b_\alpha) \, ,
\end{equation}
where $h_\pm$ are arbitrary functions on the worldsheet.
This means $e= h_+ h_- \eps^{\alpha\beta} \tau_\alpha (\pd_\beta\eta+  b_\beta)$.
Then the Nambu--Goto action reads
\begin{equation}
\begin{split}
S_{\text{NG}}[X^\mu, \eta]=T\int_{\Sigma} d^2\sigma\Big[ &-\eps^{\alpha\beta}m_\alpha(\partial_\beta \eta+ b_\beta)-\frac{1}{2}\eps^{\alpha\beta}\CB_{\alpha\beta} \\
&+\frac{\eps^{\alpha\alpha'}\eps^{\beta\beta'}\Big((\partial_{\alpha'}\eta+b_{\alpha'})(\partial_{\beta'}\eta+b_{\beta'})-\tau_{\alpha'}\tau_{\beta'}\Big)}{2\eps^{\gamma\gamma'}\tau_\gamma(\partial_{\gamma'}\eta+b_{\gamma'})}h_{\alpha\beta}\Big] \\
&+ \frac{1}{4\pi} \int_\Sigma d^2\sigma \sqrt{|\hat\gamma|} R^{(2)}(\hat\gamma) \phi\, .
\end{split}
\end{equation}
where the induced metric is
\begin{equation}
  \label{eq:gamma-hat}
\hat\gamma_{\alpha\beta}=-h_+ h_- \Big( \tau_\alpha \tau_\beta-(\pd_\alpha \eta + b_\alpha)(\pd_\beta \eta+b_\beta)\Big)\, .
\end{equation}
Note that all dependence on $h_\pm$ dropped out in the leading $\alpha' $ term.
If Weyl invariance also holds at the quantum level, it would be possible to choose $h_\pm=1$ and thus remove the conformal factor also from the dilaton term.
One could also vary the action~\eqref{eq:tnc-ns-string-action-vielbein} with respect to the vielbeine, which fixes the Lagrange multipliers.

Similarly, one can write the action with $x^M$ target space \eqref{eq:tnc-reinterpr} in Nambu--Goto form as well.
In this higher-dimensional notation, the solutions to the constraint equation~\eqref{eq:constr-solution} can be written as $e_\alpha{}^\pm= h_{\pm}(\tau_M\pm b_M)\pd_\alpha X^M$.
Then the Nambu--Goto action reads
\begin{equation}
\begin{split}
S_{\text{NG}}[X^M]=&-\frac{T}{2}\int_{\Sigma}\left[\eps^{\alpha\beta} \pd_\alpha X^M \pd_\beta X^N \bar \CB_{MN} + \sqrt{|\hat \gamma|} \hat \gamma^{\alpha\beta} \pd_\alpha X^M \pd_\beta X^N h_{MN} \right] \\
&+ \frac{1}{4\pi} \int_\Sigma d^2\sigma \sqrt{|\hat\gamma|} R^{(2)}(\hat\gamma) \phi\, .
\end{split}
\end{equation}
where $\hat\gamma_{\alpha\beta}$ is given by by~\eqref{eq:gamma-hat}.

\subsection{Gauge symmetries of the TNC string in NS-NS background}
\label{ssec:tnc-string-symmetries}
We now briefly discuss the symmetries of the TNC string action~\eqref{eq:tnc-ns-string-action-vielbein}.
Using the vielbeine~$\CE_\mu{}^a$ for the spatial metric $h_{\mu\nu}$ defined in~\eqref{eq:spatial-metric-vielbeine}, the action is invariant under  Galilean (or Milne) boosts parametrized by $\lambda_a$,
\begin{equation}
\label{eq:TNC-Gal-boost}
\delta m_\mu = \lambda_a \CE_\mu{}^a \spa \delta h_{\mu\nu}=2\tau_{(\mu}\CE_{\nu)}{}^a \lambda_a \spa \delta \lambda_\pm=-\frac{1}{e}\eps^{\alpha\beta}e_\alpha{}^{\pm} \pd_\beta X^\mu \CE_\mu{}^a \lambda_a\, ,
\end{equation}
where we used \eqref{eq:typeITNC} and wrote $\lambda_\mu = \lambda_a \CE_\mu{}^a$.

Additionally, TNC has a $U(1)$ symmetry, see for example~\cite{Duval:1984cj,Andringa:2010it,Hartong:2015zia}.
In terms of the $(d+2)$-dimensional manifold described by $x^\mM=(u,x^\mu)$ in~\eqref{eq:metric-decomposition}, it is generated by $\delta u=\sigma( x^\mu)$.
From the form of the metric in~\eqref{eq:metric-decomposition}, it follows that $m_\mu$ transforms as
\begin{equation}
  \label{eq:m-sigma-transformation}
  \delta_\sigma m_\mu = \pd_\mu \sigma\, .
\end{equation}
In addition, the magnetic field components $\CB_{\mu\nu}$ and $b_\mu$ transform as follows:
\begin{equation}
  \label{eq:bfield-sigma-transformation}
  \delta_\sigma b_\mu
    = \LL_\xi \CB_{u\mu}
    = 0\, ,
  \qquad
  \delta_\sigma  \CB_{\mu\nu}
    = \LL_\xi \CB_{\mu\nu}
    = 2 b_{[\mu} \pd_{\nu]} \sigma\, ,
\end{equation}
where $\xi=-\sigma \pd_u$.
Together, the transformations~\eqref{eq:m-sigma-transformation} and~\eqref{eq:bfield-sigma-transformation} leave the action~\eqref{eq:tnc-ns-string-action-vielbein} invariant.%
\footnote{%
  In fact, these transformations leave the Lagrangian invariant up to $\pd_\beta( \eps^{\alpha\beta}\sigma \pd_\alpha \eta)$, which is a total derivative of a function without winding and hence does not change the action.
}

Furthermore, as already remarked in~\cite{Gallegos:2019icg}, this action also has an additional $U(1)_B$ symmetry which comes from a one-form gauge transformation $\Lambda_\mM(x^\mu)=\rho(x^\mu)\delta_\mM^u$ whose only component is along $u$,
\begin{equation}
  \label{eq:tnc-u(1)-b-transf}
  \delta_\rho b_\mu = -\pd_\mu \rho\, ,
    \qquad
  \delta_\rho \eta = \rho\, .
\end{equation}
Note that the $\eta$-transformation clearly suffices to make the action invariant, but its relation to the $B$-field gauge transformation is not clear in this context.
The relation between these symmetries and the higher-dimensional interpretation of the action in terms of the $x^M$ target space is further discussed in Appendix~\ref{sapp:tnc-snc-sym-id}.

\section{Equivalence between SNC string and TNC string}
\label{sec:snc-tnc-equivalence}

In this section, we show that (under one assumption) the TNC action derived in the previous section is equivalent to the string Newton--Cartan (SNC) action of \cite{Bergshoeff:2018yvt,Gomis:2019zyu}.
More precisely, our statement is that the two theories are completely equivalent if the compact spatial longitudinal direction of the SNC geometry is an isometry.

We first illustrate how the action for closed strings in a SNC background can be obtained by taking a large-$c$ expansion of the relativistic Nambu--Goto action, extending the derivation of \cite{Andringa:2012uz}.
In the absence of a Kalb--Ramond field, we find that requiring the subleading term in the expansion of the embedding fields to drop out of the action leads to a constraint on the generalized clock form.
This constraint is closely related to the `foliation constraint' of SNC, however it can be lifted at the classical level when the $B$-field is included.
We also discuss the field redefinition freedom of the SNC string action, which plays an important role in our discussion.
After reviewing the Polyakov action for SNC strings, we specialize to the case with a longitudinal isometry and present the direct map between the SNC and TNC string actions.

\subsection{Large-\texorpdfstring{$c$}{c} expansion of relativistic Nambu--Goto action}
\label{ssec:large-c-exp}
Instead of using null reduction, a  natural approach to construct an action for non-relativistic strings is to expand the relativistic string action in a large speed of light limit. Indeed, the Gomis--Ooguri or `stringy' Newton--Cartan string was originally introduced by considering a non-relativistic limit of a string moving in Minkowski space~\cite{Gomis:2000bd,Danielsson:2000gi,Andringa:2012uz}.
The resulting action was later extended to general backgrounds~\cite{Bergshoeff:2018yvt}.

However, in the latter case, it was not possible to implement all gauge symmetries that one would expect from an algebraic point of view without implementing a \emph{foliation constraint} on the target space geometry.
While it is natural that quantum consistency of a string theory imposes restrictions on the geometry of its target space, it is somewhat surprising that such restrictions would also be necessary to formulate the theory even at a classical level.

Instead, using recent insights from the non-relativistic expansion of general relativity~\cite{Hansen:2019svu}, we will now show that it is possible to obtain an action for non-relativistic strings without imposing any restrictions on the target space geometry.
The resulting action has an accidental symmetry
 coming from the interplay between the metric and $B$-field coupling, which we can use to reproduce the SNC string action that was obtained in earlier work.

We remark that a similar accidental symmetry is present when taking
the non-relativistic limit of a charged point particle  \cite{Hansen:2019}. The resulting action for a (single) charged particle moving in TNC spacetime exhibits a shift symmetry between the NC field $m_\mu$ and the $U(1)$ gauge field $A_\mu$.
In that case, for a particle of charge $q$ and mass $m$, only the combination $q A_\mu + m\, m_\mu$ enters in the action.
A further similarity with that case is that including the coupling to the $U(1)$ gauge field lifts the constraint that $\tau$ is closed~\cite{Hansen:2019}.
We also note that the shift symmetry can be broken, for example if one would
 consider two (or more) particles that have different mass to charge ratios \cite{Festuccia:2016caf}.

\subsubsection*{Expanding the Nambu--Goto term}
In any non-relativistic limit, one must take care to cancel the divergence of the rest mass of the objects of the theory, which can be achieved using a background electromagnetic field in the case of point particles
\cite{Jensen:2014wha}. For closed strings, the same can be done using the Kalb--Ramond two-form gauge field $B_{MN}$, as in
Ref.~\cite{Gomis:2000bd,Danielsson:2000gi}.
Hence it is natural to consider non-relativistic limits of the target space geometry where two special directions are distinguished, corresponding to the orientation of the divergent electric component of the $B$-field.

However, before including the $B$-field, it will be useful to first analyze the divergencies that arise from the Nambu--Goto action in such an expansion.
Including the appropriate factors of $c$, our starting point is therefore
\begin{equation}
  \label{eq:ng-string-expansion-action-no-bfield}
  S_\text{NG}[X^M] = - Tc \int_\Sigma d^2\sigma \sqrt{-\det G_{\alpha\beta}}\, .
\end{equation}
Here, $G_{\alpha\beta}$ is the worldsheet pullback of the $D$-dimensional target space metric $G_{MN}$, which we take to be a fully general Lorentzian metric, with $M,N=0,1,\ldots D-1$.
We can now designate two `distinguished' directions by splitting $G_{MN}$ in a two-dimensional Lorentzian and a $(D-2)$-dimensional Euclidean part,
\begin{equation}
\label{Gexp}
  G_{MN}
    = c^2 \left( - E_M{}^0 E_N{}^0 + E_M{}^1 E_N{}^1\right) + \Pi_{MN}^\perp
    = - c^2 \eta_{AB} E_M{}^A E_N{}^B + \Pi_{MN}^\perp\, .
\end{equation}
Here, we have introduced tangent space indices $A=0,1$ and $A'=2,\ldots,D-1$, which are raised and lowered by $\eta_{AB}$ and $\delta_{A'B'}$, respectively. Furthermore, we impose that
\begin{equation}
E_A{}^M \Pi_{MN}^\perp = 0 \quad ,  \quad   E_A{}^M E_M{}^B  = \delta_A^B \ .
 \end{equation}
We can then write
\begin{subequations}
  \label{eq:ng-string-expansion-vielbeine}
\begin{align}
  E_M{}^A
    &= \tau_M{}^A + \frac{1}{c^2} m_M{}^A + \OOO{c^{-4}},\\
  \Pi_{MN}^\perp
    &= H_{MN}^\perp + \OOO{c^{-2}}\, .
\end{align}
\end{subequations}
which leads to the following expansion of the target space metric:
\begin{equation}
\label{eq:G-expansion}
  G_{MN} = c^2 \tau_{MN} + H_{MN} + \OOO{c^{-2}}\, .
\end{equation}
Here, we have defined
$\tau_{MN} = \eta_{AB} \tau_M{}^A \tau_N{}^B$
and
$H_{MN} = H^\perp_{MN} + 2 \eta_{AB} \tau_{(M}{}^A m_{N)}{}^B$\, .

Note that $\tau_{MN}$ is a symmetric rank two tensor, so its pullback $\tau_{\alpha\beta}$ can be used as a nondegenerate, invertible Lorentzian metric on the worldsheet.
We will use $\tau=\det{\tau_{\alpha\beta}}$, and we also denote the inverse of the induced worldsheet metric by $\tau^{\alpha\beta}$.

However, this is only possible if we consider worldsheet embeddings for which the pushforward of the worldsheet tangent vectors by the embedding map $X^M$ always have nonzero overlap with the two directions specified by the target space one-forms $\tau_M{}^A$.
Hence it is natural to only consider worldsheet embeddings that satisfy this requirement.\footnote{%
  A similar restriction applies to the relativistic string: the pullback metric on the worldsheet only has a Lorentzian signature if the worldsheet is embedded in such a way that it captures the Lorentzian signature of the target space and we should only consider embeddings that satisfy this requirement.
}
For this reason, the two directions parametrized by $\tau_M{}^A$ are often referred to as the \emph{longitudinal} directions, while the remaining $D-2$ directions are known as the \emph{transverse} directions.

To find the expansion of the Nambu--Goto action~\eqref{eq:ng-string-expansion-action-no-bfield}, we can therefore write
\begin{equation}
  G_{\alpha\beta}
    = c^2 \tau_{\alpha\gamma} \left(\delta_\beta^\gamma + \frac{1}{c^2} H^\gamma{}_\beta + \OOO{c^{-4}} \right)\, ,
\end{equation}
where $H^\gamma{}_\beta = \tau^{\gamma\delta} H_{\delta\beta}$.
Using this factorization of $G_{\alpha\beta}$, expanding the determinant and the square root in the Nambu--Goto Lagrangian leads to
\begin{align}
  \LL_\text{NG}
    &= - c^3 T \sqrt{-\tau}
      - c \frac{T}{2} \sqrt{-\tau} \tau^{\alpha\beta} H_{\alpha\beta}
      + \OOO{c\inv}\, .
\end{align}

Since the string embedding coordinates $X^M$ take values in the target space, they must also be expanded
in order to maintain diffeomorphism invariance. We thus parametrize this as
\begin{equation}
\label{eq:embedding-expansion}
  X^M = x^M + \frac{1}{c^2} y^M + \OOO{c^{-4}}\, .
\end{equation}
Following the general arguments of~\cite{Hansen:2019svu}, the subleading term parametrized by $y^M$ has the effect of entering the equations of motion of the leading order (LO) Lagrangian into the next-to-leading order (NLO) Lagrangian.
Concretely, if we write
\begin{equation}
  \LL_\text{NG}
    = c^3 \LL_\text{NG,LO} + c \LL_\text{NG,NLO} + \OOO{c^{-1}}\, ,
\end{equation}
the leading order and next-to-leading order Lagrangians are
\begin{align}
  \LL_\text{NG,LO}
  \label{eq:ng-string-expansion-lo-lag}
    &= - T \sqrt{-\tau}\, , \\
  \LL_\text{NG,NLO}
  \label{eq:ng-string-expansion-nlo-lag}
    &= - T \sqrt{-\tau} \tau^{\alpha\beta} H_{\alpha\beta}
      + y^M \frac{\delta \LL_\text{NG,LO}}{\delta x^M}\, .
\end{align}
The leading order term in the action can be cancelled using a counterterm coming from the coupling to $B_{MN}$, as we will see shortly.  We can then obtain a finite Lagrangian in the $c\to\infty$ limit by appropriately rescaling the string tension $T$.

The subleading embedding coordinates $y^M$ would in general enter in the action.
However, if the equations of motion of the leading-order action are trivial, the $y^M$-terms will not enter the action.
The leading-order part of the Lagrangian~\eqref{eq:ng-string-expansion-lo-lag} is proportional to the volume form given by the pullback $\tau_\alpha{}^A$ of the longitudinal target space vielbeine $\tau_M{}^A$.
The equations of motion of~\eqref{eq:ng-string-expansion-lo-lag} are trivial if this term is topological, which implies that the target space vielbeine have to satisfy\footnote{%
  This can be argued in more detail as follows.
  Exactness of~\eqref{eq:ng-string-expansion-lo-lag} implies that $\tau_\alpha{}^0\tau_\beta{}^1 d\sigma^\alpha\wedge d\sigma^\beta = d\alpha$ on the worldsheet.
  Denoting the embedding map from the worldsheet $\Sigma$ to the target space $M$ by $\vphi:\Sigma\to~M$ and denoting $\tau^0\wedge\tau^1$ on the target space by $\omega$, this means $d\alpha = \vphi^\ast \omega$ which implies $0=d^2\alpha = d \vphi^\ast \omega = \vphi^\ast d\omega$.
  This does not automatically mean that $d\omega=0$ on $M$, only that it vanishes on the directions that are `probed' by $\Sigma$ through the embedding $\vphi$.
    However, since the worldsheet tangent vectors always have a nonzero projection along the directions parametrized by $\tau_M{}^A$, and are otherwise arbitrary, we can infer that $d\omega=0$ in target space, which gives~\eqref{eq:simplified-foliation-constraint}.
}
\begin{equation}
  \label{eq:simplified-foliation-constraint}
  d\left(\tau^0\wedge \tau^1\right) = 0\, .
\end{equation}
To prevent the appearance of $y^M$ in the resulting non-relativistic string action, one could therefore be tempted to impose~\eqref{eq:simplified-foliation-constraint} as a constraint on the geometry on the target space, even at the classical level.
In fact, this restriction is related to the foliation constraint introduced to preserve gauge invariance in the context of the SNC string~\cite{Gomis:2019zyu,Bergshoeff:2018yvt}.
We will discuss this relation and review the appearance of the foliation constraint in the SNC theory later in Sec. \ref{ssec:snc-extensions-as-field-redef}.

However, we will now argue that after carefully taking into account the coupling to the $B$-field
at the next order in the $1/c^2$ expansion, this restriction is in fact unnecessary at the classical level.

\subsubsection*{Expanding the Wess--Zumino term}
To cancel the leading order divergence~\eqref{eq:ng-string-expansion-lo-lag} in the Nambu--Goto term, we can expand the target space Kalb--Ramond field $B_{MN}$ as follows:
\begin{equation}
\label{Bexp}
  B_{MN} = - c^2 \left(E_M{}^0 E_N{}^1 - E_M{}^1 E_N{}^0\right) + \bar B_{MN} \, .
\end{equation}
This expansion can be reproduced from an algebraic viewpoint as done in Sec. \ref{sapp:B-gauge}.
Here, the subleading component $\bar B_{MN}$ can have both longitudinal and transverse components.
It also has its own expansion in inverse powers of $c$, but we will only need the leading $c$-independent term which we denote directly with $\bar B_{MN}$.
Now note that since $\tau_{\alpha\beta}=\tau_\alpha{}^A \tau_\beta{}^B \eta_{AB}$ we can write its inverse explicitly as $\tau^{\alpha\beta}=\tau^\alpha{}_A\tau^\beta{}_B \eta^{AB}$ where we defined the inverse vectors
\begin{equation}
\label{eq:inverse-taus}
\tau^\alpha{}_0=\frac{1}{\sqrt{-\tau}}\eps^{\alpha\beta}\tau_\beta{}^1 \spa \tau^\alpha{}_1=-\frac{1}{\sqrt{-\tau}}\eps^{\alpha\beta}\tau_\beta{}^0\, .
\end{equation}
These satisfy $\tau^\alpha{}_A \tau_\alpha{}^B=\delta^A_B$\, .
Note also that $\sqrt{-\tau}=\det{\tau_\alpha{}^A}$\, .
Using the expansion of the longitudinal vielbeine in~\eqref{eq:ng-string-expansion-vielbeine}, the Wess--Zumino Lagrangian then gives
\begin{align}
  \LL_\text{WZ}
    &= - c \frac{T}{2} \eps^{\alpha\beta} B_{\alpha\beta}
    =  c^3 \LL_\text{WZ,LO} + c \LL_\text{WZ,NLO} + \OOO{c\inv},
\end{align}
where the leading and the next-to-leading terms are
\begin{align}
  \LL_\text{WZ,LO}
  \label{eq:B-string-expansion-lo-lag}
    &= T \sqrt{-\tau}\, , \\
  \LL_\text{WZ,NLO}
  \label{eq:B-string-expansion-nlo-lag}
    &= -\frac{T}{2} \sqrt{-\tau} \eps^{\alpha\beta} \bar B_{\alpha\beta}
      - T \sqrt{-\tau} \eps^{\alpha\beta} \left(\tau_\alpha{}^0 m_\beta{}^1 - \tau_\alpha{}^1 m_\beta{}^0\right)
      + y^M \frac{\delta \LL_\text{WZ,LO}}{\delta x^M}\, .
\end{align}
The leading order term indeed cancels the leading order term~\eqref{eq:ng-string-expansion-lo-lag} of the Nambu--Goto expansion.
Furthermore, we see that the terms proportional to $y^M$ in the next-to-leading order Wess--Zumino and Nambu--Goto terms \emph{also} cancel.
Consequently, it is not necessary to impose the foliation constraint~\eqref{eq:simplified-foliation-constraint} on the target space geometry at a classical level.

In addition, another remarkable cancellation takes place.
Using~\eqref{eq:inverse-taus}, we can decompose the next-to-leading Nambu--Goto term~\eqref{eq:ng-string-expansion-nlo-lag} in a transverse and a longitudinal part,
\begin{equation}
\label{eq:H-and-Hperp}
    - \frac{T}{2} \sqrt{-\tau} \tau^{\alpha\beta} H_{\alpha\beta}
    = -\frac{T}{2} \sqrt{-\tau} \tau^{\alpha\beta} H^\perp_{\alpha\beta}
      + T \sqrt{-\tau} \eps^{\alpha\beta} \left(\tau_\alpha{}^0 m_\beta{}^1 - \tau_\alpha{}^1 m_\beta{}^0\right).
\end{equation}
We then see that the longitudinal metric coupling is cancelled by the next-to-leading Wess--Zumino term in~\eqref{eq:B-string-expansion-nlo-lag}!
After rescaling $T\to T/c$ and subsequently taking $c\to\infty$, the total action is
\begin{equation}
\label{eq:SNC-NG-perp}
  S[X^M]
  = - \frac{T}{2} \int_\Sigma d^2 \sigma
    \left(
       \sqrt{-\tau} \tau^{\alpha\beta} H^\perp_{\alpha\beta}
      + \eps^{\alpha\beta} \bar B_{\alpha\beta}
    \right).
\end{equation}

Note that strings in this action only couple to the longitudinal directions through the $B$-field, since $\bar B_{MN}$ is general but $H_{MN}^\perp$ is purely transverse.
At first sight, this appears to be more restrictive than the SNC string action introduced in~\cite{Bergshoeff:2018yvt}, which allows longitudinal coupling in both the metric and the $B$-field.
However, as we will now argue, the two actions can in fact be identified due to an accidental symmetry (which already appeared in~\cite{Bergshoeff:2018yvt} as a field redefinition symmetry of the latter).
Using similar manipulations as in \eqref{eq:H-and-Hperp}, one can indeed see that
\begin{equation}
\begin{split}
 \sqrt{-\tau} \tau^{\alpha\beta} H^\perp_{\alpha\beta}
      + \sqrt{-\tau} \eps^{\alpha\beta} \bar B_{\alpha\beta}
    &= \sqrt{-\tau} \tau^{\alpha\beta} (H^\perp_{\alpha\beta}- 2 C_\alpha{}^0 \tau_\beta{}^0+ 2C_\alpha{}^1 \tau_\beta{}^1) \\
    &{}\qquad
      +\eps^{\alpha\beta}( \bar B_{\alpha\beta}-2C_\alpha{}^0 \tau_\beta{}^1+2C_\alpha{}^1 \tau_\beta{}^0)\, .
\end{split}
\end{equation}
This shows that on the level of the target space fields we have the field redefinition freedom
\begin{equation}
\label{eq:field-redef}
H^\perp_{MN} \to H^\perp_{MN} + 2 C_{(M}{}^A \tau_{N)}{}^B \eta_{AB} \spa \bar B_{MN} \to \bar B_{MN} -2 C_{[M}{}^A \tau_{N]}{}^B \epsilon_{AB}\, .
\end{equation}
Our convention for the target space tangent Levi-Civita symbol $\epsilon_{AB}$ is $\epsilon_{01}=1$.
(Note that this is opposite to our convention for $\eps_{\alpha\beta}$, see Appendix~\ref{app:conventions}.)

The choice $C_M{}^A=m_M{}^A$ then leads to
\begin{equation}
\label{eq:SNC-NG}
  S[X^M]
  = - \frac{T}{2} \int_\Sigma d^2\sigma
    \left(
       \sqrt{-\tau} \tau^{\alpha\beta} H_{\alpha\beta}
      + \eps^{\alpha\beta} B_{\alpha\beta}
    \right),
\end{equation}
where we denoted $B_{MN}=\bar B_{MN}-2 m_{[M}{}^A \tau_{N]}{}^B \epsilon_{AB}$.
This is exactly the general Nambu--Goto SNC string action introduced in~\cite{Bergshoeff:2018yvt}, excluding the dilaton term.
We have thus shown that the SNC string action can be obtained by taking the large-$c$ limit of the relativistic Nambu--Goto string action.
However, as mentioned above, by this procedure we also found that no foliation constraint is required in order for \eqref{eq:SNC-NG} to be well defined.
We will discuss the interpretation of this from the SNC point of view in the following.

\subsection{SNC Polyakov action, field redefinitions and foliation constraint}
\label{ssec:snc-extensions-as-field-redef}
The Polyakov Lagrangian for closed strings on a string Newton--Cartan (SNC) background with NS-NS fields introduced by \cite{Bergshoeff:2018yvt} is
\begin{multline}
\CL= - \frac{T}{2}\Big[ \sqrt{-\gamma}\, \gamma^{\alpha\beta}\partial_\alpha X^M \partial_\beta X^N H_{M N} +\eps^{\alpha\beta}\partial_\alpha X^M \partial_\beta X^N B_{MN}  \\
+ \lambda \eps^{\alpha\beta}  e_\alpha{}^+ \tau_M{}^+ \partial_\beta X^M + \bar\lambda\eps^{\alpha\beta}  e_\alpha{}^- \tau_M{}^- \partial_\beta X^M  \Big]
+\frac{1}{4\pi}e\CR^{(2)}\Phi\, .
\label{BGY}
\end{multline}
where $H_{MN}$ is the same as introduced below \eqref{eq:G-expansion}, and we also denote\footnote{%
  Note that instead of writing $\tau_M$ and $\bar\tau_M$ as is conventional in the SNC literature, we use~$\tau_M{}^\pm$ to avoid confusion with the clock one-form $\tau_\mu$ of TNC.
}
\begin{gather}
\tau_M{}^{\pm}=\tau_M{}^0 \pm \tau_M{}^1 \spa H^\perp_{MN}=E_M{}^{A' } E_N{}^{B' }\delta_{A' B' }\, .
\end{gather}
Integrating out the Lagrange multipliers $\lambda$ and $\bar \lambda$ in \eqref{BGY} and neglecting the dilaton term gives back the Nambu--Goto action \eqref{eq:SNC-NG}.
The background fields of SNC geometry are the generalized clock form $\tau_M {}^A$, the transverse vielbeine $E_M{}^{A' }$ and $m_M{}^A$.
In addition, the string couples to a Kalb--Ramond field $B_{MN}$ and a dilaton $\Phi$.
Here $M,N= 0,1...,D-1$ are the indices of the $D$-dimensional spacetime, while $A=0,1$ and $A' = 2,...,D-1$ are the longitudinal and transverse tangent space indices, respectively.
The two-dimensional worldsheet vielbeine are introduced as in Appendix~\ref{app:conventions}.

The field redefinition freedom \eqref{eq:field-redef} of the Nambu--Goto action is still present in the Polyakov action.
It now also transforms the Lagrange multipliers~\cite{Gomis:2019zyu,Bergshoeff:2018yvt},
\begin{subequations}
\label{BGYredef}
\begin{gather}
\label{eq:multiplier-redef}
\lambda= \lambda' +\frac{1}{e} \eps^{\alpha\beta} e_\alpha{}^- \pd_\beta X^M\bar C_M  \spa \bar\lambda= \bar\lambda' +\frac{1}{e} \eps^{\alpha\beta} e_\alpha{}^+ \pd_\beta X^M C_M\, , \\
H_{MN}=H'_{MN}+2C_{(M}{}^A \tau_{N)}{}^B \eta_{AB} \spa B_{MN} = B'_{MN}-2C_{[M}{}^A \tau_{N]}{}^B\epsilon_{AB}\, .
\end{gather}
\end{subequations}
Here, we have defined
\begin{equation}
C_M=C_M{}^0+C_M{}^1 \spa \bar C_M=C_M{}^0-C_M{}^1 \, .
\end{equation}
Again all dependence on $C_M{}^A$, $\bar C_M{}^A$ drops out in \eqref{BGY}.
As shown in~\cite{Bergshoeff:2018yvt}, the full field redefinition freedom in \eqref{BGY} is even larger, allowing for a rescaling of $\tau_M{}^A$ and the Lagrange multipliers, together with a shift in the dilaton due to the change in the path integral measure.
The parameters for this rescaling are usually denoted $C$ and $\bar C$.
However, we will not need to consider this part of the field redefinition freedom here.

The target space symmetries of SNC geometry consist of longitudinal translations and a Lorentz boost, transverse translations and rotations and string Galilean boosts.
In addition, the SNC algebra proposed in~\cite{Andringa:2012uz} contains two sets of generators that can be thought of as extensions, generally denoted by $Z_A$ and $Z_{AB}$.
We will now discuss each of these extensions in turn and show that they can be absorbed in the field redefinition \eqref{BGYredef}, which will be important for our upcoming identification of the SNC and TNC actions.

The parameter corresponding to the field transformations generated by $Z_A$ is usually denoted by $\sigma^A$.
It transforms the fields in the Lagrangian~\eqref{BGY} as
\begin{equation}
\label{eq:Z_A-transf}
  \delta m_M{}^A
    =D_M \sigma^A,
  \quad
  \delta \lambda
    =\frac{1}{e} \eps^{\alpha\beta} e_{\alpha}{}^- \pd_\beta X^M D_{M} \sigma^-,
  \quad
  \delta \bar\lambda
    =\frac{1}{e} \eps^{\alpha\beta} e_{\alpha}{}^+ \pd_\beta X^M D_{M} \sigma^+ \, .
\end{equation}
It was proposed in~\cite{Bergshoeff:2018yvt} that these transformations are only a symmetry of the SNC action \eqref{BGY} if one imposes the ``foliation constraint''
\begin{equation}
\label{Dtau_constr}
D_{[M}\tau_{N]}{}^A=0\, .
\end{equation}
Here, the covariant derivative includes the (longitudinal) spin connection $\omega_M{}^A{}_B=\omega_M \epsilon^A{}_B$.
We can relate~\eqref{Dtau_constr} to the constraint \eqref{eq:simplified-foliation-constraint} from the previous section, which is solved by
\begin{equation}
d\tau^0= A\wedge \tau^1+ C\wedge \tau^0 \spa d\tau^1=B \wedge \tau^0 - C\wedge \tau^1\, ,
\end{equation}
where $A$, $B$ and $C$ are general one-forms.
One can easily check that with the choice $A=\omega^0{}_1=\omega^1{}_0=B$ and $C=0$, the above coincides with the foliation constraint~\eqref{Dtau_constr}, where $\omega^A{}_B$ is the spin connection one-form.
In other words, solutions to the foliation constraint are a subcase of a more general class of solutions of the constraint~\eqref{eq:simplified-foliation-constraint} found when expanding the relativistic Nambu--Goto string action.
Note that this is a constraint imposed on the target space geometry already at a \emph{classical} level.
This poses an apparent contradiction with our ultimate aim of identifying SNC and TNC strings, since for the latter no restrictions on the classical geometry are necessary.

However, in the same spirit as in Sec.~\ref{ssec:large-c-exp} we now claim that in the presence of a $B$-field the constraint~\eqref{Dtau_constr} is unnecessary.
To see this, we note that one can interpret the $Z_A$ transformation \eqref{eq:Z_A-transf} as part of a larger transformation that also involves the $B$-field.
In fact, it is a specific case of the field redefinition.
Indeed, setting $C_M{}^A=-D_M \sigma^A$ in \eqref{BGYredef}, we get
\begin{subequations}
\begin{gather}
\label{ZA-redef}
m'_M{}^A=m_M{}^A+D_M \sigma^A \spa B' _{MN}= B_{MN} -2\epsilon_{AB} D_{[M} \sigma^A \tau_{N]}{}^B\, ,\\
\lambda' = \lambda +\frac{1}{e} \eps^{\alpha\beta} e_\alpha{}^- \pd_\alpha X^M D_M \sigma^- \spa \bar\lambda' = \bar\lambda +\frac{1}{e} \eps^{\alpha\beta} e_\alpha{}^+ \pd_\alpha X^M D_M \sigma^+\, .
\end{gather}
\end{subequations}
These transformations leave the action invariant without any requirement on the geometry of target space.
For this reason, we are led to think of \eqref{ZA-redef} as the `complete' $Z^A$ transformations.
Consequently, we can drop the foliation constraint \eqref{Dtau_constr}, at least at a classical level.

On the other hand, this constraint has recently been derived from a beta function calculation for SNC strings in \cite{Gomis:2019zyu}.
In the light of our findings above, this suggests that the foliation constraint can be viewed as an equation of motion of the corresponding non-relativistic theory of gravity, in the same way that Einstein's equations emerge from the beta functions of relativistic string theory.\footnote{%
  On the TNC side, the combination of \eqref{Dtau_constr} and the upcoming \eqref{tau-H} leads to
  \begin{equation}
  \label{eq:foliation-implication}
  \pd_{[\mu} \tau_{\nu]}=0 \spa \pd_{[\mu} b_{\nu]}=0\, .
  \end{equation}
  This follows from the fact that the string-NC longitudinal spin connection $\omega_M^{AB}$ is zero in our gauge, as can be easily verified putting $M=\mu$ and $N=v$ in \eqref{Dtau_constr}.
  As remarked above, these conditions only need to hold at the quantum level.
  In fact, also TNC target spaces with $d\tau\neq 0$ should be allowed, as they can be obtained from Penrose limits of $AdS_5\times S^5$ as discussed in \cite{Harmark:2018cdl}.
  In that case, RR fields should also be included, which would change the beta functions.
}

Likewise, transformations coming from the second extension generator $Z_{AB}$ can also be identified as part of the field redefinitions.
Parametrizing the action of $Z_{AB}$ using~$\sigma_{AB}$, one of the resulting field transformation is $\delta m_M{}^A=\sigma^A{}_B \tau_M{}^B$.
This transformation can be reproduced by setting $C_M{}^A=\sigma^A{}_B \tau_M{}^B$ in \eqref{BGYredef}.
For general~$\sigma_{AB}$, the resulting transformation also affects the Lagrange multipliers and the $B$-field.
To understand this symmetry from an algebraic perspective, it will likely be easiest to restrict to $\eta^{AB}\sigma_{AB}=0$, which leaves the $B$-field invariant.

Having shown that all the extension symmetries are part of the larger field redefinition, it follows that we no longer need to consider them once we have fixed the field redefinition freedom.
This is essentially what we need to identify the general SNC action~\eqref{BGY} with the action~\eqref{eq:SNC-NG}, eliminating $m_M{}^A$ from the description.
The importance of this will be even clearer in the next section.
Indeed, mapping the SNC string action to the TNC string action is very easy after fixing the redundancy due to the field redefinitions.

\subsection{Longitudinal spatial isometry and map to TNC string}
\label{ssec:map-from-snc-with-isometry-to-tnc}
In the original work \cite{Gomis:2000bd,Danielsson:2000gi}, where the flat space version of the SNC string action was first derived, it was already pointed out that such strings only have a nontrivial spectrum if they have nonzero winding along a spatial longitudinal direction of the SNC target space.
This is also taken to hold for strings in general SNC backgrounds~\cite{Gomis:2019zyu}.
Hence SNC strings should be studied on a target space with a compact longitudinal direction.
In the following, we will add the assumption that there is an isometry along this direction.

Choosing coordinates such that $\pd_v$ parametrizes the compact spatial isometry, we can split the space-time indices as $M=(v,\mu)$.
None of the fields in the worldsheet action will depend on $X^v$.
Since $v$ is longitudinal and spatial we can choose coordinates such that $\tau_v {}^0 =0$, $\tau_v {}^1 \neq 0$ and $E_v {}^{A'}=0$.
Then we can always choose a gauge in which\footnote{%
  This can be achieved through a longitudinal Lorentz transformation $\delta \tau_M{}^A=\Lambda \epsilon^A{}_B \tau_M{}^B$.
}
\begin{equation}
\label{gauge}
\tau_v{}^0=0\, , \qquad \tau_v{}^1=1\, , \qquad E_v{}^{A' }=0\, .
\end{equation}

Starting from the general action~\eqref{BGY}, we want to reduce to the form~\eqref{eq:SNC-NG} where only the fields $H^\perp_{\alpha\beta}$ and $\bar B_{\alpha\beta}$ enter.
In the Polyakov formulation, this can be done by making use of \eqref{id1} and \eqref{id2}, which allow us to rewrite \eqref{BGY} as
\begin{align}
\label{eq:BGY1}
\CL = &-\frac{T}{2}\left[e\gamma^{\alpha\beta} H^\perp_{\alpha\beta}+\eps^{\alpha\beta}\left(B_{\alpha\beta}-2\tau_\alpha{}^1 m_\beta{}^0+2\tau_\alpha{}^0 m_\beta{}^1\right)\right]
  \\\nonumber
      &+Te\left[\lambda-\frac{1}{e}\eps^{\alpha\beta} e_\alpha{}^-\left(m_\beta{}^0-m_\beta{}^1\right)\right]e^{\alpha'}{}_- \tau_{\alpha'}{}^+
  \\\nonumber
      &-Te\left[\bar\lambda-\frac{1}{e} \eps^{\alpha\beta} e_\alpha{}^+ \left(m_\beta{}^0+m_\beta{}^1\right)\right]e^{\alpha'}{}_+\tau_{\alpha'}{}^-\, .
\end{align}
From the form of the action in \eqref{eq:BGY1} it is now manifest that we can completely remove the $m_\alpha{}^A$ fields by redefining the $B$-field and the Lagrange multipliers.\footnote{%
The fact that electric coupling allows one to remove the $m$-field also occurs for a point particle, where the Maxwell one-form combines with the Newton--Cartan one-form  $m_\mu$. See e.g. Section 6 of \cite{Festuccia:2016caf}.
}
Equivalently, this can be achieved using the field redefinitions \eqref{BGYredef} by setting $C_M{}^A=m_M{}^A$.
In other words, we can always choose a gauge such that $m'_M{}^A=0$.
Reinstating the dilaton term, we then get
\begin{equation}
\label{eq:BGY2}
\CL = -\frac{T}{2}\left[e\gamma^{\alpha\beta} H^\perp_{\alpha\beta}+\eps^{\alpha\beta}\bar B_{\alpha\beta}+ \lambda'\eps^{\alpha\beta} e_\alpha{}^+ \tau_\beta{}^+
+  \bar\lambda' \eps^{\alpha\beta} e_\alpha{}^-\tau_\beta{}^- \right]
+\frac{1}{4\pi}e\CR^{(2)}\Phi\, .
\end{equation}
The action \eqref{eq:BGY2} is still invariant under string Galilean boosts, even though it is no longer written explicitly in terms of the boost-invariant combination $H_{MN}$.
This is due to the fact that the new Kalb--Ramond field $\bar B_{MN}$ and Lagrange multipliers $\lambda'$, $\bar\lambda'$ transform under a Galilean boost.
We will discuss the symmetries of this action in more detail in Appendix~\ref{app:SNC-TNC-symmetries}.
Integrating out the Lagrange multipliers gets us back to the Nambu--Goto type action \eqref{eq:SNC-NG}.

We are now ready to present the map between SNC and TNC strings.
This map was already discussed in a restricted setting in Appendix A of \cite{Harmark:2018cdl} which did not yet include the full NS-NS coupling on the TNC side.
Having now coupled the TNC strings to the full NS-NS sector in Section~\ref{sec:null-red}, we can present the complete picture.

Comparing the SNC action \eqref{eq:BGY2} to the higher-dimensional interpretation of the TNC action in~\eqref{eq:tnc-reinterpr}, we see that the two are equivalent if we identify
\begin{subequations}
\label{eq:map-of-backgrounds}
\begin{gather}
X^v=\eta \spa \tau_\mu{}^0=\tau_\mu \spa  \tau_\mu{}^1=b_\mu \spa H^\perp_{\mu\nu} = h_{\mu\nu}\, , \label{tau-H}\\
\bar B_{MN} = \bar \CB_{MN} \spa \Phi=\phi \label{B}\, ,\\
\lambda' =\lambda_+  \spa \bar\lambda'=\lambda_- \, .
\end{gather}
\end{subequations}
Here we have used the gauge choice~\eqref{gauge}.
This is a one-to-one map
between the string action and background fields of SNC with a compact longitudinal spatial isometry
and the TNC string action.%
\footnote{%
  The identification can also be made without eliminating $m_M{}^A$, i.e. in terms of the original $B_{MN}$ instead of $\bar B_{MN}$.
In that case one would have
$$ \bar \CB_{MN}=B_{MN} + (m_M{}^A \tau_N{}^B-m_N{}^A \tau_M{}^B) \epsilon_{AB}\, ,$$
which, using the fact that $\bar \CB_{v\mu}=-m_\mu$, can be translated into
$$m_\mu=m_\mu{}^0-B_{v\mu}+m_v{}^1 \tau_\mu{}^0-m_v{}^0 \tau_\mu{}^1 \spa \CB_{\mu\nu}=B_{\mu\nu}+2B_{v[\mu}\tau_{\nu]}{}^1-2m_v{}^1\tau_{[\mu}{}^0\tau_{\nu]}{}^1-2m_{[\mu}{}^1 \tau_{\nu]}{}^0\, .$$
The other identifications are unchanged, and equivalence of the actions still holds.
However, notice that in this way the map would no longer be one-to-one, but all SNC backgrounds related by a field redefinition correspond to the same TNC background.
}

A discussion of this result is in order.
First, the compact longitudinal spatial direction~$v$ above indeed corresponds to the $v$-direction of the higher-dimensional interpretation of the TNC string action that was discussed in Section~\ref{ssec:tnc-action-uplift-vielbein-form}.
Second, the symmetries of both actions are identified, as we discuss in more detail in Appendix~\ref{sapp:tnc-snc-sym-id}.
As pointed out above, the SNC action was originally formulated with a constraint on the generalized clock form, which restricted the allowed target space geometries even at the classical level.
Such a restriction is unnecessary and even unwanted from the point of view of the TNC action, and we discussed the resolution of this apparent tension in Section~\ref{ssec:snc-extensions-as-field-redef}.

In other words, in this section we have shown that (provided the compact longitudinal direction is an isometry) the non-relativistic strings obtained from a null-reduction procedure in Section~\ref{sec:null-red} are equivalent to the non-relativistic strings resulting from a large-$c$ expansion as derived in Section~\ref{ssec:large-c-exp}.

Furthermore, it is interesting to note that the identifications in~\eqref{eq:map-of-backgrounds} coincide with what one would get from the longitudinal spatial T-duality map of SNC to a Lorentzian geometry with a null isometry\footnote{%
See equation (3.12) in~\cite{Bergshoeff:2018yvt} with $\tau_{yy}=1$ and $\tau_{yi}=\tau_i{}^1$, $\tilde y=u$, $x^i=x^\mu$.
There, $y$ has to be identified with our $v$.}
if we identify the Lorentzian metric of this geometry with $G_{\mM\mN}$ in \eqref{eq:metric-decomposition}.

Thus, our procedure and results can also be thought of as giving a different viewpoint on the notion of a null T-duality, relating relativistic strings in a NS-NS background with a null isometry to non-relativistic strings on a background with a longitudinal spatial isometry.
This also connects to recent work \cite{Kluson:2019qgj} where this kind of T-duality was investigated from a Hamiltonian perspective.
However, the connection of our results to \cite{Bergshoeff:2018yvt} is still to be clarified.
There, the null isometry direction is assumed to be compact, whereas in our derivation the non-compactness of $u$ played a crucial role.
Thus, this is an aspect that needs to be studied in more detail.

\section{Strings with non-relativistic worldsheet in NS-NS background}
\label{sec:nr-ws-limit}
In the preceding sections we have studied strings propagating in non-relativistic target spaces.
However, so far the worldsheet geometry has been relativistic.
By taking a suitable limit of the target space geometry whilst simultaneously rescaling the worldsheet fields, one can obtain a string action that describes non-relativistic geometry both on the worldsheet and in the target space~\cite{Harmark:2018cdl}.
The resulting strings are equipped with an infinite-dimensional symmetry algebra on the worldsheet, known as the Galilean Conformal Algebra (GCA).
For particular backgrounds, these strings have been shown~\cite{Harmark:2017rpg,Harmark:2018cdl} to reproduce the continuum version of Spin Matrix Theory limits of $\mN=4$ SYM theory~\cite{Harmark:2014mpa}, thus leading to a covariant string description of (for example) the $SU(2)$ Landau-Lifshitz model.

So far, the $B$-field and dilaton were not included in this non-relativistic worldsheet limit.
We will now briefly discuss an extension of the limit as applied to the Polyakov action~\eqref{eq:tnc-ns-string-action-vielbein}, including the $B$-field and dilaton coupling.
We first focus on the Nambu--Goto and $B$-field coupling in Section~\ref{ssec:nr-ws-theory-leading} and then study the behaviour of the dilaton term in Section~\ref{ssec:nr-ws-dilaton}.

\subsection{Non-relativistic worldsheet sigma-model with Kalb--Ramond field}
\label{ssec:nr-ws-theory-leading}
Following~\cite{Harmark:2018cdl}, we now obtain a theory of strings with non-relativistic worldsheet geometry using a large $c$ limit.
In contrast to the $1/c$ expansion discussed in Section~\ref{ssec:large-c-exp}, which only concerned the target space fields and the string embedding maps, this limit will act on all target space and worldsheet fields.
Our starting point is the Lagrangian of~\eqref{eq:tnc-ns-string-action-vielbein}, which at leading order in $\alpha'$ is given by
\begin{equation}
  \begin{split}
    \LL_\text{leading}
    &= - \frac{T}{2}
      \left[
        2\eps^{\alpha\beta} m_\alpha \left(\pd_\beta \eta + b_\beta \right)
          + e \eta^{ab} e^\alpha{}_a e^\beta{}_b h_{\alpha\beta}
          + \eps^{ab} \CB_{\alpha\beta}
      \right]
      \\
    &{}\qquad
      -T\eps^{\alpha\beta} \left[
          \lambda_+ e_\alpha{}^+
            \left(\tau_\beta + \pd_\beta\eta + b_\beta\right)
          + \lambda_- e_\alpha{}^-
            \left(\tau_\beta - \pd_\beta\eta - b_\beta\right)
        \right].
  \end{split}
\end{equation}
Following~\cite{Harmark:2017rpg}, we scale the tension and the target space fields by
\begin{equation}
  \label{eq:tnc-target-space-variables-scaling}
  T \to \frac{ T}{c}
  \spa
  \tau_\mu \to c^2 \tau_\mu
  \spa
  \eta \to c\, \eta
  \spa
  m_\mu \to m_\mu
  \spa
  h_{\mu\nu} \to h_{\mu\nu}\, .
\end{equation}
The embedding maps $X^\mu(\sigma^\alpha)$ are not rescaled.
To retain the coupling to $\CB_{\mu\nu}$ and $b_\mu$ in the large $c$ limit of the action, we scale them by\footnote{%
  Said differently, the higher-dimensional Kalb--Ramond field $\CB_{\mM \mN}$ (which leads to $\CB_{\mu\nu}$ and $b_\mu$ via the null reduction discussed in Section~\ref{ssec:nsns-tnc-action-from-null-reduction}) has a uniform scaling $\CB_{\mM \mN} \to c\, \CB_{\mM \mN}$.
}
\begin{equation}
  \CB_{\mu\nu} \to c\, \CB_{\mu\nu}
  \spa
  b_\mu \to c\, b_\mu\, .
\end{equation}
Finally, if we parametrize the Lagrange multipliers using $\lambda_{\pm} = (\omega\pm\psi)/2$, we obtain a finite limit of the constraints by rescaling
\begin{subequations}
  \label{eq:worldsheet-fields-scaling}
\begin{alignat}{3}
  \label{eq:vielbeine-scaling}
  &e^0_\alpha \to c^2\, e^0_\alpha
    \spa
  &&e^1_\alpha \to c\, e^1_\alpha\, ,
  \\
  \label{eq:lagrange-multipliers-scaling}
 & \omega \to \frac{\omega}{c^3}
    \spa
  &&\psi \to \frac{\psi}{c^2}\, .
\end{alignat}
\end{subequations}
This results in the following action for strings with a non-relativistic worldsheet geometry,
\begin{equation}
  \label{eq:leading-alpha-prime-limit-action}
  \begin{split}
    S_\text{leading}
      = - \frac{T}{2} \int d^2 \sigma
      &\left[
          2 \eps^{\alpha\beta} m_\alpha \left(\pd_\beta \eta + b_\beta\right)
          + e e^\alpha{}_1 e^\beta{}_1 h_{\alpha\beta}
          + \eps^{\alpha\beta} B_{\alpha\beta}
        \right.\\
      &{}\qquad\left.
        +\omega \eps^{\alpha\beta} e_\alpha{}^0 \tau_\beta
        +\psi \eps^{\alpha\beta} \Big(e_\alpha{}^1 \tau_\beta + e_\alpha{}^0 \left(\pd_\beta \eta + b_\beta\right)\Big)
      \right].
  \end{split}
\end{equation}
We recall that the target-space geometry after the limit becomes $U(1)$-Galilean~\cite{Harmark:2017rpg},
which is the geometry that corresponds to gauging a spacetime symmetry algebra consisting of a direct sum
of the (massless) Galilean algebra and a $U(1)$ factor.
Moreover, the local symmetries \eqref{eq:m-sigma-transformation}, \eqref{eq:tnc-u(1)-b-transf} and \eqref{eq:tnc-u(1)-b-transf} still leave \eqref{eq:leading-alpha-prime-limit-action} invariant.
This means the parameter $\rho$ for the $U(1)_B$ symmetry also needs to scale as $\rho \to c\rho$, whereas $\sigma$ does not scale as known already from~\cite{Harmark:2018cdl}.
So both $U(1)$ symmetries survive in the limit
This action is furthermore invariant under the local transformations
\begin{equation}
e_\alpha {}^0 \rightarrow f   e_\alpha {}^0 \spa e_\alpha {}^1\rightarrow f  e_\alpha {}^1  + \hat{f}  e_\alpha {}^0 \spa \omega \rightarrow \frac{1}{f} \omega - \frac{\hat{f}}{f^2} \psi \spa \psi \rightarrow \frac{1}{f}\psi\,,
\end{equation}
exactly as in the case without $B$-field \cite{Harmark:2018cdl}.
These symmetries correspond to local Weyl transformations ($\hat f=0$) and local Galilean boosts ($f=1$), which can be used to obtain flat worldsheet vielbeine $e_\alpha{}^a=\delta_\alpha^a$.
Furthermore, in this gauge it is possible to see that the action has a residual Galilean conformal symmetry whose classical generators satisfy
\begin{equation}
  [L_m,L_n] = (m-n) L_{m+n},
    \qquad
  [L_m,M_n] = (m-n) M_{m+n},
\end{equation}
which is known as the Galilean conformal algebra (GCA).
This algebra has appeared previously in tensionless limits of relativistic strings, see for example \cite{Isberg:1993av} and also \cite{Bagchi:2009my}.

\subsection{Dilaton term and a criterion for non-relativistic worldsheets}
\label{ssec:nr-ws-dilaton}
So far, the dilaton coupling has been unaffected by our operations in Sections~\ref{sec:null-red} and~\ref{sec:snc-tnc-equivalence}.
In the action~\eqref{eq:tnc-ns-string-action-vielbein} it is still given in terms of the two-dimensional (Lorentzian) scalar curvature,
\begin{equation}
  \label{eq:dilaton-term-separate}
  S_\text{dil} = \frac{1}{4\pi} \int_\Sigma d^2\sigma \sqrt{|\gamma|} R^{(2)} \phi \, .
\end{equation}
As is apparent from Equation~\eqref{eq:worldsheet-fields-scaling}, the limit we consider acts differently on timelike and spacelike worldsheet vielbeine and we therefore take a Lorentzian worldsheet geometry as the starting point of our discussion.

We assume that the dilaton itself does not scale under the worldsheet and target space scalings in~\eqref{eq:tnc-target-space-variables-scaling} and~\eqref{eq:worldsheet-fields-scaling}.
This is natural, since any scaling of the relativistic string coupling $g_s = e^\phi$ would only shift the dilaton.
Thus we only need to consider the effect of the scaling limit on the curvature of the worldsheet.

Given a set of worldsheet vielbeine $e^a=e_\alpha{}^a d\sigma^\alpha$, we can solve the 2D Cartan structure equation
to obtain the associated spin connection
\begin{equation}
  \omega
    = \omega_\alpha d\sigma^\alpha
    = 2 \eps^{ab} e^\beta{}_b \pd_{[\alpha} e_{\beta]a} d\sigma^\alpha
    = \eps^{ab} \LL_{e_a} \theta_b \, .
\end{equation}
Here, we have defined the one-forms $\theta_a= \eta_{ab} e^b$.
We can then write the dilaton coupling~\eqref{eq:dilaton-term-separate} using the curvature $d\omega$ of this spin connection as follows,
\begin{equation}
  S_\text{dil}
    = \frac{1}{2\pi} \int_\Sigma \phi \, d\omega \, .
\end{equation}
Note that the scaling in~\eqref{eq:worldsheet-fields-scaling} should not be interpreted as a coordinate transformation on the worldsheet, not even at finite values of $c$.
For that reason, the vielbein scaling~\eqref{eq:vielbeine-scaling} only affects the curvature $d\omega$ through the way it depends on $e^a$ and $\theta_a$ and not through any derivatives.
Dropping all tildes, this means
\begin{equation}
  S_\text{dil}
  \to \frac{1}{2\pi} \int_\Sigma \phi\, d\left( c \LL_{e_1} \theta_0 + \frac{1}{c} \LL_{e_0} \theta_1\right) \, .
\end{equation}

Since we do not rescale $\phi$, this presents us with a criterion that the \emph{worldsheet} geometry has to satisfy in order to survive the non-relativistic limit $c\to\infty$.
For simplicity, here we restrict ourselves to worldsheets without punctures.
Assuming the resulting non-relativistic sigma model should be well defined for a generic dilaton profile, we see that the dilaton coupling diverges unless
\begin{equation}
\label{eq:restriction-on-geometries}
  d \left( \LL_{e_1} \theta_0\right) = 0 \, .
\end{equation}
Relativistic worldsheet geometries for which this does not hold have a divergent action as $c\to\infty$ and will therefore be suppressed in the path integral.

In fact, this condition fits well with the non-relativistic interpretation of the worldsheet geometry defined by $e^0$ and $e^1$ after the limit.
One obvious solution to~\eqref{eq:restriction-on-geometries} is requiring $e^0$ (and hence $\theta_0$) to be exact, which means there exists an absolute time on the worldsheet.
This holds for example for an infinite cylinder, where
\begin{equation}
  e^0 = dT \, ,
    \quad
  e^1 = d\vphi \, ,
    \qquad
  T\in(-\infty,\infty) \, ,
    \quad
  \vphi\in[0,2\pi) \, .
\end{equation}
Similarly, a flat torus has $d\omega=0$ and is therefore also allowed.
This suggests that the connection between these non-relativistic sigma models and the continuum limit of non-relativistic quantum-mechanical spin chains of Spin Matrix Theory, which was already established on the level of the classical action~\cite{Harmark:2017rpg}, could even extend to the dynamics of the resulting strings.
At least locally, these strings have a notion of absolute time on the worldsheet, and the transverse worldsheet direction could then be seen as the continuum spin chain.
To make this relation more precise, it would be very interesting to include punctures in this discussion and to understand the effective coupling constant of the resulting string theory.
We hope to come back to these issues in the near future.

\section{Discussion}

We conclude by discussing some open problems and outlining further interesting directions.

The main result of this paper is the equivalence between the TNC and SNC actions for non-relativistic strings.
As remarked in the above, our map between the theories requires that the compact longitudinal direction of the SNC string is an isometry.
It would be interesting to examine whether this is a restriction or in some way inherent in the formulation of the SNC string itself.

In other treatments of the SNC string, a `foliation constraint' is usually imposed on the target space geometry, but we find that this is not necessary if part of the target space symmetries are implemented differently (by also acting on the Kalb--Ramond $B$-field).
The SNC string theory we identify with the TNC theory is therefore more general, but still contains the solutions that satisfy the foliation constraint.

At this point our discussion is entirely classical.
However, it is likely that some form of constraints on the target space geometry will arise in the quantum theory, following the recent work on beta functions in non-relativistic string theories~\cite{Gomis:2019zyu,Gallegos:2019icg}.
Note that the beta functions of the TNC theory have not yet been explored in a fully general Kalb--Ramond background.
It would then be possible to study if our identification of SNC and TNC non-relativistic string theories still holds on a quantum level.

Our procedure of deriving the TNC action relies on doing a reduction on a non-compact null direction
and subsequently going to a sector of fixed momentum in the null direction. The way this is implemented
is reminiscent of the Ro\v{c}ek-Verlinde approach to T-duality~\cite{Rocek:1991ps}, but differs in the details.
It would be good
to connect our construction to the Hamiltonian interpretations of T-duality along a null direction put forward in \cite{Kluson:2018egd,Kluson:2018vfd,Kluson:2019qgj}.
It would also be useful to further analyze T-duality transformations of non-relativistic strings when there is a compact spatial isometry
in the transverse space (see  also \cite{Bergshoeff:2018yvt}).

The inclusion of worldsheet fermions in non-relativistic string theory, the addition of
worldsheet and spacetime supersymmetry  as well as the role of RR background fields are further obvious directions to consider.\footnote{%
  See the recent work~\cite{Blair:2019qwi} for progress in this direction.
}
Moreover, a careful analysis of the open string sector of non-relativistic string theory and the notion of non-relativistic D-branes%
\footnote{See \cite{Kluson:2017vwp,Kluson:2017djw,Kluson:2017abm,Kluson:2019ifd} for Hamiltonian
analyses related to non-relativistic branes.}
also span important open problems.
There have also been recent studies of integrability and classical solutions in the context of non-relativistic string theory \cite{Roychowdhury:2019vzh,Roychowdhury:2019olt}.
Finally, it seems likely that there are further useful connections to doubled geometry
\cite{Ko:2015rha,Morand:2017fnv,Berman:2019izh}.

Turning to the non-relativistic worldsheet models discussed in Section \ref{sec:nr-ws-limit}, there are also
numerous open issues to investigate.  First of all, a more detailed analysis of the allowed worldsheet
geometries, also including punctures, will be important to understand the effective coupling constant of the resulting string theory. In particular, there could be a fascinating underlying relation to interactions in SMT
\cite{Harmark:2008gm,Harmark:2014mpa}.%
\footnote{See also \cite{Harmark:2016cjq} for the $SU(2)$ SMT limit of the non-abelian Born-Infeld action for D-branes in the AdS/CFT correspondence.}
The SMT limit is connected to a limit of AdS/CFT, and thus relies on the presence of RR fields. This is another
motivation to better understand RR fields in non-relativistic string theory, even before applying the worldsheet limit.
Related to this, it would also be interesting to see what the $B$-field corresponds to in the context of SMT, now that
we know how to include it in the non-relativistic worldsheet sigma model.

\section*{Acknowledgements}

We thank Eric Bergshoeff and Ziqi Yan for useful discussions.
The work of JH is supported by the Royal Society University Research Fellowship ``Non-Lorentzian Geometry in Holography'' (grant number UF160197).
The work of TH and NO is supported in part by the project ``Towards a deeper understanding of  black holes with non-relativistic holography'' of the Independent Research Fund Denmark (grant number DFF-6108-00340).
The work of GO and NO is supported in part by Villum Foundation Experiment project 00023086.
LM thanks Niels Bohr Institute, Perugia University and INFN of Perugia for support.

\appendix
\section{Conventions and identities for worldsheet vielbeine}
\label{app:conventions}

Here we collect the conventions for the worldsheet vielbeine and review some useful identities.
The two-dimensional vielbeine (or zweibeine) are introduced through
\begin{equation}
\gamma_{\alpha\beta}=e_\alpha{}^a e_\beta{}^b \eta_{ab}\, .
\end{equation}
Our conventions for Lévi--Civita symbols are $\eps^{01} = -\eps_{01} =+1$ for both the worldsheet symbol $\eps^{\alpha\beta}$ and the tangent space symbol $\eps^{ab}$.
Note that this is opposite to the conventional definition for the longitudinal target space symbol $\epsilon_{AB}$, where $\epsilon_{01}=+1$, so we use different letters to distinguish the two.
The vielbeine determinant is
\begin{equation}
e = \det ( e_\alpha {}^a ) = e_\tau {}^0 e_\sigma {}^1 - e_\tau {}^1 e_\sigma {}^0\, .
\end{equation}
Using this we can compute $\eps_{ab} e_\alpha {}^a e_\beta {}^b = e \, \eps_{\alpha\beta}$.
The inverse vielbeine are given by
\begin{equation}
e^\alpha {}_a = \frac{1}{e} \eps^{\alpha\beta} e_\beta {}^b \eps_{ba}\, ,
\end{equation}
and they obey
$e_\alpha {}^a e^\beta{}_a = \delta_\alpha ^\beta$ and $e_\alpha {}^a e^\alpha{}_b = \delta^a_b$.
Moreover, we find
\begin{gather}
\frac{1}{e} = \det ( e^\alpha {}_a ) = e^0 {}_0 e^1 {}_1 - e^0 {}_1 e^1 {}_0\,,
  \qquad
\eps^{ab} e^\alpha {}_a e^\beta {}_b = \frac{1}{e}  \eps^{\alpha\beta}\, .
\end{gather}
At various places in the main text, it is useful to consider the null combinations
\begin{equation}
e_\alpha {}^\pm = e_\alpha {}^0 \pm e_\alpha {}^1\, .
\end{equation}
The inverses are then given by as $e^\alpha {}_\pm = (e^\alpha {}_0 \pm e^\alpha {}_1)/2$.
They satisfy
\begin{gather}
e_\alpha {}^+ e_\beta {}^- = - \gamma_{\alpha\beta} + e\, \eps_{\alpha\beta}
\spa
e_\alpha {}^- e_\beta {}^+ = - \gamma_{\alpha\beta} - e\, \eps_{\alpha\beta}\, , \nonumber
\\
e^\alpha{}_+ e_\alpha {}^+ = e^\alpha{}_- e_\alpha {}^- =1 \spa e^\alpha{}_+ e_\alpha {}^- = e^\alpha{}_- e_\alpha {}^+ =0\, ,
\\
e^\alpha {}_+ e^\beta {}_- = - \frac{1}{4}\gamma^{\alpha\beta} - \frac{1}{4e} \eps^{\alpha\beta}
\spa
e^\alpha {}_- e^\beta {}_+ = - \frac{1}{4}\gamma^{\alpha\beta} + \frac{1}{4e} \eps^{\alpha\beta}\, . \nonumber
\end{gather}
From this, we can obtain
\begin{align}
\frac{1}{e}\eps^{\alpha \alpha' } \eps^{\beta \beta' }(e^-_{\alpha' } e^+_{\beta' } +e_{\alpha' }^+ e_{\beta' }^- ) &= 2 e \gamma^{\alpha \beta} \label{id1}\, ,  \\
\frac{1}{e}\eps^{\alpha \alpha' } \eps^{\beta \beta' } (e^-_{\alpha' } e^+_{\beta' } - e_{\alpha' }^+ e_{\beta' }^- )&= 2\eps^{\alpha \beta} \label{id2}\, .
\end{align}
These identities are crucial to the discussion in Section~\ref{sec:snc-tnc-equivalence}.

\section{Review of SNC symmetries and identification with TNC symmetries}
\label{app:SNC-TNC-symmetries}
For completeness, we first briefly review `stringy' Newton--Cartan (SNC) geometry and how the
associated symmetry algebra  introduced in~\cite{Andringa:2012uz} can be obtained from
a $1/c$ expansion of the Poincar\'e  algebra.
This complements our discussion of the
expansion of the relativistic Nambu--Goto action in \eqref{Gexp}.
We also include some observations on how one may include the gauge transformation of the NS-NS $B$-field in the resulting algebraic structure.
Finally, following the identification of the SNC and TNC string actions in Section~\ref{sec:snc-tnc-equivalence}, we discuss how the gauge symmetries of the torsional Newton--Cartan (TNC) and SNC strings can be related explicitly.

\subsection{SNC symmetry algebra}
\label{sapp:SNC-algebra}

A $D$-dimensional SNC geometry has the property that it is decomposed into two  \emph{longitudinal} directions
and  $D-2$ \emph{transverse} directions.
The geometry along each of the directions can be parametrized using two separate sets of vielbeine
\begin{equation}
  \tau_M{}^A \, ,
    \qquad
  E_M{}^{A'} \, .
\end{equation}
Here, $A=0,1$ are flat indices on the longitudinal directions, while $A'=2,\ldots,D-2$ correspond to the transverse directions.
Our convention for the longitudinal Lévi-Civita symbol $\epsilon_{AB}$ is $\epsilon_{01}=+1$.
(Note that we use $\eps^{01}=+1$ for the worldsheet symbol $\eps^{\alpha\beta}$ and the worldsheet tangent space symbol $\eps^{ab}$.)

The local symmetries of SNC geometry include transverse translations and rotations $P_{A'}$ and $M_{A'B'}$, longitudinal translations and rotations $H_{A}$ and $M_{AB}=\epsilon_{AB}M$ and the so-called `string' Galilean boosts $G_{AB'}$.
These can be seen to arise from the Poincar\'e generators $M_{\ua\ub}$, $P_\ua$ ($\ua,\ub = 0 \ldots D-1$)
in $D$ dimensions as follows.
Splitting up $\ua=(A,A')$ in terms of the longitudinal and transverse directions
defined above, one identifies
$M_{AB'} = c G_{AB'}$, $P_A = H_A/c$ and the remaining generators unchanged.
Then, using the method of Lie algebra expansions%
\footnote{See for example Refs.~\cite{deAzcarraga:2002xi,Izaurieta:2006zz,Khasanov:2011jr}
and also the recent applications \cite{Hansen:2019vqf,Bergshoeff:2019ctr,Gomis:2019fdh}.}
the following graded Lie algebra is obtained (by tensoring with the polynomial ring in the variable $\sigma = 1/c^2$):
\begin{equation}
[ M_{AB}^{(m)}, H_C^{(n)} ]= 2 \eta_{C [A } H_{B]}^{(m+n)} \spa
[ M_{A'B'}^{(m)}, P_{C'}^{(n)} ]=2 \delta_{C '[A' } P_{B']}^{(m+n)} \, ,
\label{SNCLie}
  \end{equation}
  \begin{equation}
 [ H_{A}^{(m)}, G_{BA'} ^{(n)} ]= - \eta_{A B } P_{A'}^{(m+n)} \spa
 [ P_{A'}^{(m)}, G_{AB'} ^{(n)} ]=  \delta_{A' B' } H_{A}^{(m+n+1)}  \, ,
  \end{equation}
  \begin{equation}
[ G_{AA'}^{(m)}, G_{BB'} ^{(n)} ]= \eta_{A B } M_{A'B'}^{(m+n+1)} + \delta_{A' B' } M_{AB}^{(m+n+1)}\, ,
\end{equation}
\begin{equation}
[ G_{A A '}^{(m)}, M_{BC}^{(n)} ]=  - 2\eta_{A [B } G_{C]A'}^{(m+n)} \spa
[ G_{A A '}^{(m)}, M_{B'C'}^{(n)} ]=  -2 \delta_{A' [B' |} G_{A | C']}^{(m+n)} \,,
\end{equation}
\begin{equation}
[ M_{A' B'}^{(m)}, M_{C'D'}^{(n)} ]= 4 \delta_{[A' [C' } M_{B' ] D ']}^{(m+n)}  \spa
[ M_{AB}^{(m)}, M_{CD}^{(n)} ] = 0 \, ,
\end{equation}
where we note that  $M_{AB}^{(m)} = M^{(m)}\epsilon_{AB}$.
In this infinite dimensional Lie algebra we can consistently set to zero all the generators with grade $m \geq 2$ since they form
an ideal. Furthermore, a consistent further truncation is to only keep the level 0 generators (for which
we drop the superscript from now on) supplemented
with $Z_A \equiv H_A^{(1)}$ along with $Z_{AB} \equiv M_{AB}^{(1)}$.%
\footnote{This is analogous to the way in which the Bargmann algebra can be obtained from the Poincar\'e algebra using Lie algebra expansion \cite{Khasanov:2011jr}, where in that case one decomposes with only one longitudinal direction (i.e. time) and the transverse ones being the spatial directions.}

This leads to the SNC algebra \cite{Andringa:2012uz}  with commutation relations
\begin{alignat}{3}
  [M_{A'B'}, M_{C'D'}] &= 4\delta_{[A'[C'} M_{B']D']} \, ,   \label{SNCalg1}
    &\qquad\qquad
  [M_{A'B'}, P_{C'}] &= 2\delta_{C'[A'} P_{B']} \, ,
    \\
  [M_{A'B'}, G_{CD'}] &= 2\delta_{D'[A'|} G_{C|B']} \, ,
    &\qquad\qquad
  [M_{AB}, G_{CD'}] &= 2\eta_{C[A} G_{B]D'} \, ,
    \\
  [M_{AB}, H_C] &= 2\eta_{C[A} H_{B]} \, ,
    &\qquad\qquad
  [G_{AB'}, H_C] &= \eta_{AC} P_{B'} \, .
\end{alignat}
extended with the generators $Z_{A}$ and $Z_{AB}$ satisfying
\begin{alignat}{3}
  [P_{A'}, G_{BC'}] &= \delta_{A'C'} Z_B \, ,
    &\qquad\qquad
  [M_{AB}, Z_C] &= 2\eta_{C[A} Z_{B]} \, ,
    \\
  [G_{AB'}, G_{CD'}] &= \delta_{B'D'} Z_{AC} \, ,
    &\qquad\qquad
  [Z_{AB},H_C] &= 2\eta_{C[A} H_{B]} \, .    \label{SNCalg2}
\end{alignat}
 Another way to obtain this algebra through an expansion was presented in \cite{Bergshoeff:2019ctr}, which also reproduces the  non-relativistic algebras of \cite{Ozdemir:2019orp}.

\subsection{Gauge transformations of the NS-NS \texorpdfstring{$B$}{B}-field}
\label{sapp:B-gauge}
We now introduce a useful algebraic perspective on the gauge symmetries of the $B$-field, which further motivates its expansion in the Nambu--Goto action as discussed in Section~\ref{ssec:large-c-exp}.

First, we consider the inclusion of the $B$-field in the algebra of gauge symmetries of the relativistic string.
We denote the full tangent space index by $\ua$, which we will later split in longitudinal and transverse components $\ua=(A, A' )$.
Inspired by double field theory \cite{Hull:2009mi,Hohm:2010pp}, a natural starting point is to add
a second set of ``translations'' $Q_\ua$ to the Poincar\'e algebra, transforming as Lorentz vectors under $M_{\ua\ub}$
and commuting with ordinary translations $P_\ua$. Using the fact that Lorentzian geometry can
be obtained by gauging the Poincar\'e algebra and turning local translations into diffeomorphisms we now
extend this by including the $Q_\ua$ generator in the gauging.

We write the gauge field as
\begin{equation}
  A_M= e_M{}^\ua P_\ua + \frac{1}{2} \omega_M{}^{\ua\ub} M_{\ua\ub} + \pi_M{}^\ua Q_\ua\, ,
\end{equation}
where the new element is the gauge connection  $\pi_M{}^\ua$ of the $Q_\ua$ generators. This transforms
under gauge transformations as
\begin{equation}
\delta_\Lambda \pi_M{}^\ua=\pd_M \kappa^{\ua} + \sigma^{\ua}{}_\ub \pi_M{}^\ub-\tensor{\omega}{_M^\ua_\ub}\kappa^\ub\, ,
\quad  \Lambda = \zeta^\ua P_\ua + \frac{1}{2} \sigma^{\ua\ub} M_{\ua\ub}+\kappa^\ua Q_\ua\, .
\end{equation}
In addition to translations and Lorentz transformations there now also is a parameter~$\kappa^\ua$
corresponding to the ``doubled'' translations.  Writing
\begin{equation}
  \Lambda = \xi^M A_M +\Sigma \, ,  \quad  \Sigma=\frac{1}{2} M_{\ua\ub} \lambda^{\ua\ub}+ k^\ua Q_\ua\, , \quad \lambda^{\ua\ub}=\sigma^{\ua\ub}-\omega_M{}^{\ua\ub} \xi^M\, , \quad k^\ua= \kappa^\ua-\pi_M{}^\ua \xi^M\, ,
\end{equation}
one finds that under the $\bar \delta$ variation defined in \cite{Hartong:2015zia} we have
\begin{equation}
\label{dbar-pi}
\bar\delta \pi_M{}^\ua=\LL_\xi \pi_M{}^\ua+\pd_M k^\ua+\lambda^\ua{}_\ub \pi_M{}^\ub-\tensor{\omega}{_M^\ua_\ub}k^\ub\, .
\end{equation}

Now define the following object
\begin{equation}
\label{B-def}
  B_{MN} =  e_{[M}{}^\ua \pi_{N]}{}^\ub \eta_{\ua\ub} \, .
\end{equation}
Using \eqref{dbar-pi} along with the corresponding transformation  $\bar\delta e_{M}{}^\ua$ as well as  the torsion constraint  $R(P)_{MN}^\ua =0$, this can be shown to transform as
\begin{equation}
\bar \delta B_{MN}= \LL_\xi B_{MN} + 2\pd_{[M}\alpha_{N]} \, ,
\end{equation}
where we have defined  $\alpha_M \equiv -\frac{1}{2}e_M{}^\ua k_\ua$.
In fact, it is not necessary to impose the torsion constraint if one further modifies the variation $\bar \delta$.
Thus indeed $B_{M N}$ is a (0,2) tensor
transforming under one-form gauge transformations.

This algebraic perspective on the $B$-field can be integrated into the SNC algebra we discussed in \ref{sapp:SNC-algebra} as follows.
First, we split $P_\ua$ and $Q_\ua$ as
\begin{equation}
  P_A = \frac{1}{2}\left(\frac{1}{c}H_A + c K_A\right)
    \spa
  Q_A = \frac{1}{2}\epsilon_A{}^B \left(\frac{1}{c} H_B - c K_B \right)
    \spa
  P_{A'}
    \spa
  Q_{A'}\, .
\end{equation}
This leads to a redefinition of the corresponding gauge fields.
If we parametrize
\begin{equation}
  e_M{}^\ua P_\ua + \pi_M{}^\ua Q_\ua
  = E_M{}^{A} H_A + \Pi_M{}^A K_A + e_M{}^{A'} P_{A'} + \pi_M{}^{A'} Q_{A'}\, ,
\end{equation}
the new gauge fields will be given by
\begin{equation}
  E_M{}^A = \frac{1}{2c} \left(e_M{}^A - \epsilon^A{}_B \pi_M{}^B\right)
  \spa
  \Pi_M{}^A = \frac{c}{2} \left(e_M{}^A + \epsilon^A{}_B \pi_M{}^B\right) .
\end{equation}
As a result, the $B$-field in~\eqref{B-def} now gives
\begin{gather}
  B_{MN}
    = - c^2 \epsilon_{AB} E_{[M}{}^A E_{N]}{}^C
        + \bar{B}_{MN}\, ,
  \\
  \bar{B}_{MN}
    = 2\epsilon_{AB} E_{[M}{}^A \Pi_{N]}{}^B
    + \delta_{A'B'} e_{[M}{}^{A'} \pi_{N]}{}^{B'}
    + \OOO{c^{-2}}.
\end{gather}
Thus, we have retrieved the expansion of the $B$-field that was introduced in~\eqref{Bexp}.
Note that the subleading term $\bar B_{MN}$ contains both transverse and longitudinal components.

\subsection{Identification of TNC and SNC symmetries}
\label{sapp:tnc-snc-sym-id}

In Section~\ref{ssec:map-from-snc-with-isometry-to-tnc}, we have identified the SNC string action (on backgrounds where the compact spatial longitudinal direction is an isometry) with the string action obtained from null reduction in Section~\ref{sec:null-red}.
As we discussed in Section~\ref{ssec:snc-extensions-as-field-redef}, the symmetries corresponding to the extension generators $Z_A$ and $Z_{AB}$ can be absorbed in the accidental symmetry of the string action.
We will now show explicitly how the TNC symmetries associated to the latter can be mapped to the remaining SNC gauge symmetries.

First, since the transverse spatial directions of SNC and TNC coincide, the symmetries along these directions can be equated immediately.
Following the identification presented in Section~\ref{ssec:map-from-snc-with-isometry-to-tnc}, the longitudinal Lorentz symmetry of SNC has been fixed with the gauge choice
\begin{equation}
\tau_v{}^0=0\, , \qquad \tau_v{}^1=1\, , \qquad E_v{}^{A' }=0\, .
\end{equation}
Longitudinal time translations also agree since we have identified the clock one-forms $\tau_\mu=\tau_\mu{}^0$.
On the other hand, it is not immediately obvious how coordinate transformations along the SNC direction $v$ should be interpreted in terms of the TNC theory.
Likewise, we will also have to explain what happens to the string Galilean boosts generated by $G_{1 A' }$.

Let us first look at a general coordinate transformation acting on the SNC geometry, generated by $\xi$
We have chosen coordinates such that $\pd_v$ is a Killing vector field and we want $\xi$ to preserve this property, which implies $\pd_v \xi^M=0$.
If we also require that the gauge choice $E_v{}^{A' }=0$ is preserved, we have
\begin{equation}
0=\delta E_v{}^{A' }=\CL_\xi E_{v}{}^{A' }+\lambda^{A' }{}_A \tau_v{}^A= \lambda^{A' }{}_1\, ,
\end{equation}
hence our gauge choice fixes the string Galilean boosts $G_{1A'}$ to zero.

The remaining SNC Galilean boosts $G_{0A'}$ can be identified with the TNC Galilean boosts.
Explicitly, the transverse vielbeine transform under $G_{0A'}$ as $\delta E_\mu{}^{A' }=\lambda^{A' }{}_{0}\tau_\mu{}^0$.
Note that using $H^\perp_{\mu\nu}=h_{\mu\nu}$ (which we established in the last equation of \eqref{tau-H}) we can identify
\begin{equation}
E_\mu{}^{A' }=\CE_\mu{}^{a}\, .
\end{equation}
Here, we identify the transverse tangent space indices $A'=2,\ldots,d+1$ with the spatial TNC tangent space indices $a=1,\ldots,d$ introduced in~\eqref{eq:spatial-metric-vielbeine}.
Comparing with the TNC transformations under Galilean boosts in \eqref{eq:TNC-Gal-boost}, we then see that $\lambda^0{}_{A' }= \lambda_a$ as expected.
In fact, the TNC Galilean boosts also transform the Lagrange multipliers.\footnote{This can also be avoided if one makes explicit use of the constraint imposed by the multipliers themselves.}
This is correctly reproduced if we keep track of the transformations for the redefined SNC fields in the Lagrangian \eqref{eq:BGY2},
\begin{subequations}%
\label{eq:new-multiplier-transform}%
\begin{gather}
\label{eq:induced-B}
\delta \bar B_{v\mu}=-\lambda^0{}_{A' } E_\mu{}^{A' } \spa \delta \bar B_{\mu\nu}=2 \lambda^0{}_{A' }E_{[\mu}{}^{A' } \tau_{\nu]}{}^1 \, , \\
\delta \lambda' = -\frac{1}{e}\eps^{\alpha\beta} e_\alpha{}^- \pd_\beta X^\mu E_\mu{}^{A' } \lambda^0{}_{A' } \, ,  \\
\delta \bar\lambda' = -\frac{1}{e}\eps^{\alpha\beta} e_\alpha{}^+ \pd_\beta X^\mu E_\mu{}^{A' } \lambda^0{}_{A' } \, .
\end{gather}
\end{subequations}
Note that these transformations indeed ensure that the SNC action~\eqref{eq:BGY2} is invariant under the remaining SNC Galilean boosts.
Through the identifications \eqref{eq:map-of-backgrounds}, the transformations in~\eqref{eq:new-multiplier-transform} precisely reproduce the correct transformations for $\CE_\mu{}^a$, $m_\mu$ and the Lagrange multipliers $\lambda_{\pm}$, whereas they leave $\CB_{\mu\nu}$ and $b_\mu$ invariant.

The $U(1)$ symmetry of TNC that we discussed in Section~\ref{ssec:tnc-string-symmetries} is related to the one-form gauge transformations of the SNC $B$-field $B_{MN}$.
In fact, we can reproduce the $U(1)$ transformations~\eqref{eq:bfield-sigma-transformation} using $A_M= \sigma \delta_M^v$,
keeping in mind the identifications $\bar B_{v\mu}=-m_\mu$ and $\bar B_{\mu\nu}=\bar \CB_{\mu\nu}$.
Hence we see that the TNC $U(1)$ gauge symmetry corresponding to diffeomorphisms along the null isometry direction has a natural interpretation in SNC as a gauge transformation of the $B$-field along the longitudinal compact spatial isometry.

Nicely enough, this situation is exactly mirrored for the $U(1)_B$ symmetry given in~\eqref{eq:tnc-u(1)-b-transf}.
As discussed in Section~\ref{ssec:tnc-string-symmetries}, it corresponds on the TNC side to gauge transformations of the Kalb--Ramond field along the null isometry direction $u$, together with a shift in $\eta$.
This now has a very natural interpretation on the SNC side as a shift $\delta X^v=\delta \eta=\rho$, coming from a coordinate transformation $\delta v =\xi^v=\rho$ which also leads to the transformation
\begin{equation}
  \delta b_\mu = \LL_\xi \tau_\mu{}^1= -\pd_\mu \rho\, .
\end{equation}
This point of view also explains why $\eta$ is charged under $U(1)_B$, which was not obvious in the $(d+1)$-dimensional TNC description of Section~\ref{ssec:tnc-string-symmetries}.
This completes our identification of the symmetries of the SNC and TNC string actions.

\providecommand{\href}[2]{#2}\begingroup\raggedright\endgroup

\end{document}